\documentclass[fleqn,usenatbib]{mnras}

\usepackage{newtxtext,newtxmath}

\usepackage[T1]{fontenc}
\usepackage{ae,aecompl}

\usepackage{graphicx}	
\usepackage{amsmath}	
\usepackage{amssymb}	

\title[Mass Functions from Clustering Redshifts]{Mass Functions, Luminosity Functions, and Completeness Measurements from Clustering Redshifts}

\author[D. J. Bates et al.]{
Dominic J. Bates,$^{1}$\thanks{E-mail: db217@st-andrews.ac.uk (KTS)}
Rita Tojeiro,$^{1}$
Jeffrey A. Newman,$^{2}$
Violeta Gonzalez-Perez,$^{3}$
\newauthor
Johan Comparat,$^{4}$
Donald P. Schneider,$^{5,6}$
Marcos Lima,$^{7}$
Alina Streblyanska$^{8}$
\\
$^{1}$School of Physics and Astronomy, University of St Andrews, North Haugh, St Andrews KY16 9SS, UK\\
$^{2}$Department of Physics and Astronomy, University of Pittsburgh, 3941 O'Hara Street, Pittsburgh, PA 15260\\
$^{3}$University of Portsmouth Institute of Cosmology \& Gravitation, Dennis Sciama Building, Portsmouth, PO1 3FX, UK\\
$^{4}$Max-Planck-Institut für Extraterrestrische Physik
Giessenbachstraße, 85748 Garching, Germany\\
$^{5}$Department of Astronomy and Astrophysics, The Pennsylvania State University, University Park, PA 16802\\
$^{6}$Institute for Gravitation and the Cosmos, The Pennsylvania State University,University Park, PA 16802\\
$^{7}$Departamento de F\'isica Matem\'atica, Instituto de F\'isica, Universidade de S\~ao Paulo, CP 66318, CEP 05314-970, S\~ao Paulo, SP, Brazil\\
$^{8}$Instituto de Astrofısica de Canarias
38205 La Laguna, Tenerife, Spain\\
}

\date{Accepted XXX. Received YYY; in original form ZZZ}

\pubyear{2017}

\begin{document}
\label{firstpage}
\pagerange{\pageref{firstpage}--\pageref{lastpage}}
\maketitle

\begin{abstract}
This paper presents stellar mass functions and i-band luminosity functions for Sloan Digital Sky Survey (SDSS) galaxies with $i < 21$ using clustering redshifts. From these measurements we also compute targeting completeness measurements for the Baryon Oscillation Spectroscopic Survey (BOSS). Clustering redshifts is a method of obtaining the redshift distribution of a sample of galaxies with only photometric information by measuring the angular crosscorrelation with a spectroscopic sample in different redshift bins. We construct a spectroscopic sample containing data from the BOSS + eBOSS surveys, allowing us to recover redshift distributions from photometric data out to $z\simeq 2.5$. We produce k-corrected i-band luminosity functions and stellar mass functions by applying clustering redshifts to SDSS DR8 galaxies in small bins of colour and magnitude. There is little evolution in the mass function between $0.2 < z < 0.8$, implying the most massive galaxies form most of their mass before $z = 0.8$. These mass functions are used to produce stellar mass completeness estimates for the Baryon Oscillation Spectroscopic Survey (BOSS), giving a stellar mass completeness of $80\%$ above $M_{\star} > 10^{11.4}$ between $0.2 < z < 0.7$, with completeness falling significantly at redshifts higher than 0.7, and at lower masses. Large photometric datasets will be available in the near future (DECaLS, DES, Euclid), so this, and similar techniques will become increasingly useful in order to fully utilise this data.
\end{abstract}




\section{Introduction}
\label{sec:intro}
\setlength{\parskip}{0pt} 

Large spectroscopic galaxy surveys are extremely useful tools for studying galaxy evolution. They allow us to determine stellar masses, star formation histories, and dynamics for large numbers of galaxies. In particular, deep, small area surveys such as PRIMUS \citep{coil11}, DEEP2 \citep{newm13}, and VIPERS \citep{guzz14} contain data for galaxies over a broad range of masses, colours, morphologies, and redshifts, allowing tests of galaxy evolution on very different objects.

These surveys, however, are not ideal for investigating galaxy evolution at the highest masses, since the number density of galaxies above, for example, $M_{\star} > 10^{11.5} M_{\odot}$, is extremely low. Due to their small area, these pencil-beam surveys typically only target tens or hundreds of galaxies above this mass, and are strongly affected by sample variance.

An ideal approach to study galaxy evolution at these masses is using large-volume cosmological redshift surveys, which typically target the highest mass galaxies over very large regions of the sky. The Baryon Oscillation Spectroscopic Survey (BOSS) \citep{eise11,daws13} is the most extensive of these to date, measuring spectra for roughly $1.5$ million luminous red galaxies (LRGs) over 10,000 deg$^{2}$ of sky at $z < 0.7$. BOSS contains over 100,000 galaxies with stellar masses $M_{\star} > 10^{11.5} M_{\odot}$ \citep{mara13}, so it is able to study this end of the mass function with very little shot noise. Ongoing and future surveys such as eBOSS \citep{blan17,daws16} and DESI \citep{desi16} will extend this study to higher redshifts and larger numbers of galaxies, providing additional data to better probe these masses.

One limitation of these surveys, however, is that they are optimised for cosmology, not galaxy science. Their target selection therefore involves a number of complex colour cuts, leading to samples of galaxies that are incomplete in both stellar mass and colour. To study galaxy evolution at these masses, we must quantify this incompleteness.

One method of determining incompleteness is by comparing the distribution of galaxies as a function of mass in one sample, to that of another sample which is complete in stellar mass. In \cite{leau16}, they chararacterise the stellar mass completeness of BOSS using Stripe 82, a narrow region of the SDSS with deeper $ugriz$ photometry, as well as near-IR photometry from the UKIRT Infrared Deep Sky Survey (UKIDSS) \citep{lawr07}, allowing for more accurate photometric redshifts and stellar masses. 


Large-area broad-band photometric surveys such as the Sloan Digital Sky Survey (SDSS) \citep{york00} are complete, provide data over a large area ($\simeq 14000 $ deg$^{2}$) for galaxies over a range of magnitudes and colours, so would be ideal for this purpose. One disadvantage, however, is that for SDSS-like data, photometric redshifts can be unreliable \citep{rahm16b}. In this paper, we outline a method of computing luminosity and mass functions (and hence completeness) from broad-band surveys using a technique known as clustering redshifts.

Clustering redshifts is a method of obtaining the redshift distribution of set of galaxies via crosscorrelation with a spectroscopic sample \citep{newm08}. A number of slightly different techniques have been proposed; however, the main idea is the same: the two-point angular crosscorrelation is measured between a photometric sample of galaxies and different redshift bins of a spectroscopic sample. If the photometric sample overlaps in redshift with a particular bin of the spectroscopic sample, then the measured crosscorrelation will have a positive amplitude. Combining this crosscorrelation with bias information of the two samples, it is possible to accurately measure the redshift distribution of a photometric sample of galaxies.

In this paper we use clustering redshifts to recover the redshift distributions of samples of galaxies from the SDSS photometric survey in small bins of magnitude and colour, isolating galaxies of similar type. After recovering redshift distributions of bins, we use these colours and redshifts to compute stellar masses and luminosities by examining simulated galaxies in the same bins of colour-redshift space. Finally, we compute targeting completeness for the BOSS spectroscopic sample.

The layout of this paper is as follows: Section \ref{sec:data} describes both the real and mock data used in this study. Section \ref{sec:clustz} presents the clustering redshifts method and bias correction, and our method of computing stellar masses. In section \ref{sec:testmeth2} we test our clustering redshifts method on mock data, and determine how accurately mass and luminosity functions can be recovered using this technique. In section \ref{sec:mlfunc}, this technique is applied to real SDSS photometry to produce real mass and luminosity functions. Section \ref{sec:comp} presents completeness measurements for BOSS using these computed mass and luminosity functions. Finally, in section \ref{sec:conc}, we discuss these results, and outline possible extensions of this work.

\section{Data}
\label{sec:data} 

This study uses data from two main sources. We seek to compute mass functions from photometric data. In order to perform this task, we apply the clustering redshifts technique, which requires both an ``unknown sample'' (i.e. a photometric sample of unknown redshifts) and a ``reference sample'' (a sample over the same region of sky but with spectroscopic redshifts). The unknown sample is crosscorrelated with the reference sample to recover the redshift distribution.

\subsection{BOSS \& eBOSS Reference Sample}
\label{sec:data_ref}

As a reference sample, we use data from both the SDSS-III: BOSS \citep{eise11,daws13,gunn06,smee13} and SDSS-IV: eBOSS \citep{blan17,daws16,gunn06,smee13} surveys. BOSS is a cosmological redshift survey that has measured spectra for $\sim 1.5$ million luminous red galaxies (LRGs) out to $z=0.7$, over roughly $10,000$ deg$^{2}$ of sky. Its primary aim is to map the spatial distribution of the highest mass galaxies over large volumes in order to measure the scale of baryon acoustic oscillations (BAO) in the clustering of galaxies. eBOSS is the extension of this program, ongoing at the moment, targeting $\sim 375,000$ LRGs \citep{prak16} at $0.7 < z < 0.8$, and $\sim 740,000$ quasars \citep{myer15} over the range $0.9 < z < 3.5$ both over $7,500$ deg$^{2}$ of sky. 

\begin{figure}
	\includegraphics[width=\columnwidth]{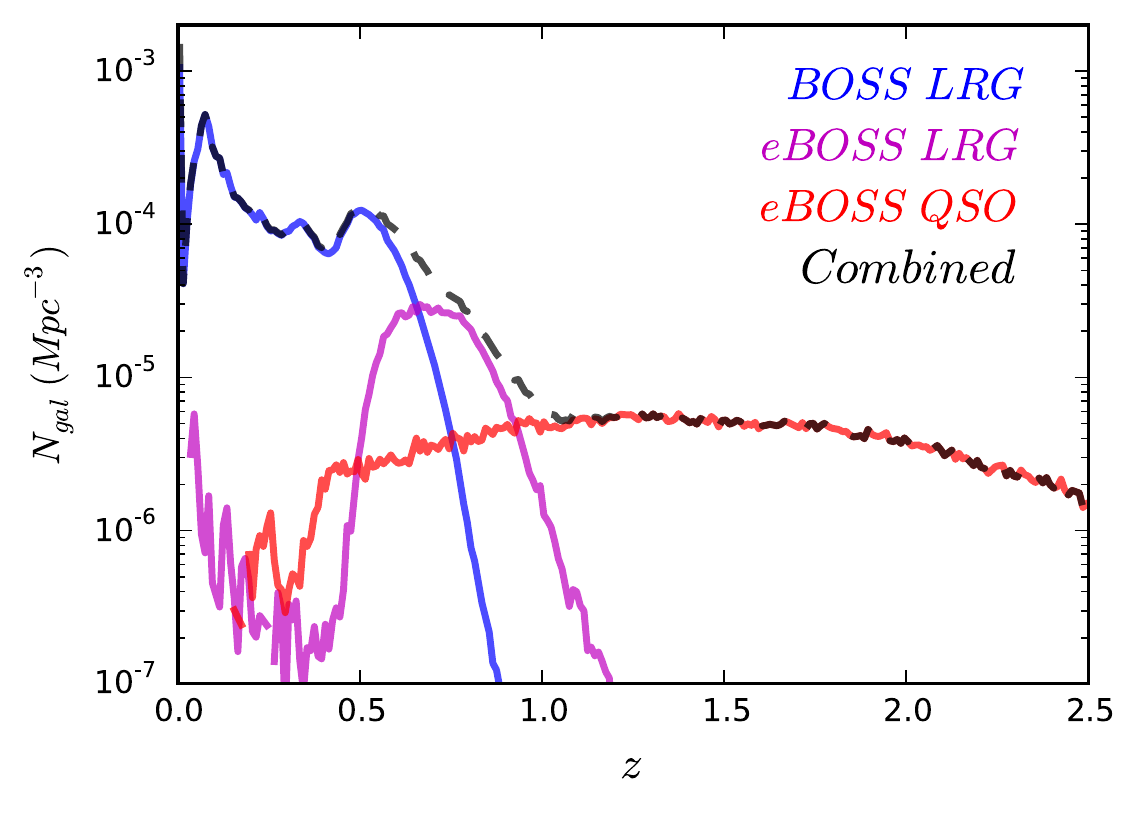}
    \caption{The comoving number density of three different spectroscopic galaxy samples described in section \ref{sec:data_ref}: At low redshift, BOSS DR12 LRGs (blue), intermediate redshifts, eBOSS DR14 LRGs (magenta), and higher redshifts, eBOSS DR14 quasars (red). The number density for the combination of these three samples is shown as the black dotted line.}
    \label{fig:number_dens}
\end{figure}

These samples are ideal for a reference sample, as they cover a large area and provide continuous, large numbers of redshifts over the range $0 < z < 3$. The reference sample is therefore a combination of BOSS DR12 \citep{alam15} LRGs in the North Galactic Cap (NGC) covering 6851 deg$^2$ and eBOSS DR14 data \citep{abol18}, containing both the LRG and quasar samples, covering 1011 and 1214 deg$^2$ respectively, within the BOSS NGC area. The total sample is then 1.1 million galaxies. The number density of both the individual and combined samples are shown in figure \ref{fig:number_dens}. When computing correlation functions in later sections, we use large scale structure catalogues from \cite{reid12} for the BOSS LRG sample, \cite{baut17} for eBOSS LRGs, and \cite{ata18} for eBOSS quasars, using 10x randoms for all samples.

\subsection{SDSS Photometric Survey}
\label{sec:data_phot}

Our photometric survey (i.e., the sample for which we wish to compute redshift distributions, along with masses and luminosities) consists of data from the SDSS photometric survey \citep{york00,gunn98}. We use photometry from DR8 \citep{aiha11} which contains $u$, $g$, $r$, $i$ and $z$ band information. We select only objects morphologically classified as galaxies, and only use data from the primary survey (i.e., the best observation for each object). To create our catalogues, we use $g$, $r$ and $i$ band modelMag magnitudes (see \cite{stou02}). We also constrain the sample to $i<21$ to avoid incompleteness, and only include galaxies in the same region as the BOSS NGC DR12 footprint using the following masks detailed in \cite{ande12}: The survey geometry mask, and veto masks for bright stars, unphotometric seeing and bright objects. Finally, we remove all galaxies that are also in our reference sample, leaving 53 million galaxies over $\sim$ 7000 deg$^2$. We create random catalogues for this sample using the MANGLE software \citep{swan08}, using 10x randoms.


\subsection{Mock Surveys}
\label{sec:data_mock}

In Section \ref{sec:testmeth2}, we assess the reliability of our method on mock data, which requires both a mock reference sample and photometric survey. In later sections, we also use semi-analytic models to compute masses from colours and redshifts. Both these purposes require mock samples, and hence lightcones from two semi-analytic models (SAMs). We firsly take data from LGalaxies \citep{henr15}. This lightcone covers 1/8th of the sky and is run on the Millennium simulation \citep{spri05} rescaled to Planck cosmology \citep{plan14}. Magnitudes are computed using \cite{mara11} stellar population models (SSPs), using a \cite{krou01} initial mass function (IMF). We also use a smaller lightcone from SAGE \citep{crot16} covering (100 deg$^2$ Area), run on the MultiDark MDPL2 simulation \citep{klyp16,kneb18}, with SEDs and magnitudes also computed using \cite{mara11} SSPs and a \cite{krou01} IMF. Both catalogues have angular positions, redshifts, SDSS magnitudes (apparent and absolute) with reddening applied \citep{calz00}, and present day stellar masses. 

We add photometric errors to the magnitudes of both SAMs by looking at how the error on a fitted magnitude in the SDSS varies as a function of that magnitude (i.e. $g$ vs $\sigma g$, $r$ vs $\sigma r$, $i$ vs $\sigma i$). For every mock galaxy, we use its magnitude to compute the mean error at this magnitude in SDSS, then draw a Gaussian random error using this value as the standard deviation. We compute errors for all mock galaxies in the $g$, $r$ and $i$ bands, and add these errors to our mock galaxy magnitudes.

From these simulated galaxy catalogues, we define a mock reference sample and photometric survey. We define our reference sample by applying the colour and magnitude cuts of the BOSS survey described in \cite{daws13} to both our LGalaxies and SAGE catalogues. This procedure produces samples with comparable redshift distribution to the BOSS survey and is further discussed in section \ref{sec:test_mf_lf}. We refer to these samples as $BOSS_{LGalaxies}$ and $BOSS_{SAGE}$. To create mock SDSS photometric samples, we cut both catalogues to $i < 21$ as in section \ref{sec:data_phot}, and also remove all galaxies present in our mock reference sample. We refer to these mock photometric surveys as $SDSS_{LGalaxies}$ and $SDSS_{SAGE}$.

\section{Method}
\label{sec:clustz} 

Crosscorrelations have long been used to test for physical association \citep{seld76}; however, the idea of using crosscorrelations to produce accurate redshift distributions has only become common over the last decade, partly due to the increase in data from large volume spectroscopic and photometric surveys. 

\cite{phil85} investigated determining correlation functions from samples with only partial redshift information; later, in \cite{phil87}, luminosity functions are computed given the assumption that galaxies close in the sky are likely at the same redshift. \cite{schn06} more generally investigate this technique by measuring crosscorrelations with galaxies binned by photometric redshift. This approach is built on more formally in \cite{newm08}, and later \cite{matt10} and \cite{matt12}, where a method is outlined for computing redshift distributions by measuring the angular crosscorrelation between a photometric sample and different redshift bins of a spectroscopic sample. The amplitude of the crosscorrelation is fitted by an analytical form; since the redshift distribution inferred also depends on the evolution in bias of both samples, an iterative technique is employed to correct for this, assuming that clustering amplitude is proportional in both the spectroscopic and photometric sample.

Some variants on this method have subsequently appeared. For example, \cite{schm13} and \cite{mena13}, propose a similar technique, measuring angular crosscorrelations with a spectroscopic sample, but over constant physical scale. Furthermore, bias evolution is corrected for by assuming a bias evolution law, and the effect of this assumption is tested, down to non-linear scales. More recent studies applying these methods include \cite{rahm15}, \cite{rahm16a}, \cite{rahm16b}, \cite{scot16}, and \cite{scot18}. \cite{vand18} present a model for computing luminosity functions using clustering information and apparent magnitudes.

In \cite{gatt18} the performance of three of these methods are investigated: \cite{newm08}, \cite{schm13} and \cite{mena13}. They apply all methods to simulated Dark Energy Survey (DES) data, finding that \cite{newm08} method produces slightly noisier redshift distributions due to having two extra degrees of freedom when fitting the crosscorrelation amplitude; furthermore, they report that the proportional bias assumption is not always accurate. 

In our preliminary tests, the noise of all techniques was largely due to noise in the crosscorrelation functions; the choice of method made only small differences to the noisiness of the recovered $n(z)s$. The main difference between methods is how the bias evolution correction is applied. 

Since, firstly, \cite{gatt18} find that Menard method appears to produce slightly less noisy distributions, and secondly, we will be investigating methods of correcting for bias, we choose to adopt a method based on \cite{mena13}.

\subsection{Clustering Redshifts Methodology}
\label{sec:clustz_meth}

The method is detailed in \cite{mena13} (hereon M13); we summarise the important points here, along with our alterations. The method is centered around computing the crosscorrelation between a photometric or ``unknown'' sample, and a number of redshift bins of a spectroscopic or ``reference'' sample.

Using the simplest \cite{peeb74} estimator, the angular crosscorrelation between two samples, 1 and 2, can be defined as, $\omega_{12}(\theta) = D_1D_2(\theta)/R_1R_2(\theta) - 1$, where $D_1D_2(\theta)$ is the number of galaxies in sample 1 separated by an angular distance $\theta$ from galaxies in sample 2. $R_1R_2(\theta)$ is the same statistic, but instead for two purely randomly distributed sets of points. The crosscorrelation function therefore describes, as a function of angle, the excess probability that galaxies in one sample will be situated a particular distance from galaxies in another. If the two samples considered overlap in redshift, they will occupy the same density field, and their positions will be correlated, hence this crosscorrelation will have a positive amplitude.

To produce an $n(z)$ measurement, we therefore need to measure the angular crosscorrelation, $\omega_{ur}(\theta,z)$, between an unknown sample, and different redshift bins of a reference sample. Since we are interested in how the amplitude of this quantity evolves with redshift, we integrate over $\theta$ to produce 

\begin{equation}
\bar{\omega}_{ur}(z) = \int_{\theta_{min}}^{\theta_{max}} d\theta W(\theta) \omega_{ur}(\theta,z)
\label{eq:int_cross}
\end{equation}

where $W(\theta)$ is the weight function, $W(\theta) = \theta^{-1}$, designed to optimise the signal-to-noise ratio. In order to probe the same physical scale at all redshifts, $\theta_{min}$ and $\theta_{max}$ are computed differently for each redshift, such that they correspond the same physical scales $r_{p,min}$ and $r_{p,max}$.

From M13, the integrated crosscorrelation is,

\begin{equation}
\bar{\omega}_{ur}(z) \propto \frac{dN_{u}}{dz}(z) \bar{b}_u(z) \bar{b}_r(z) \bar{\omega}_{DM}(z)
\label{eq:nz1}
\end{equation}

where $\frac{dN_{u}}{dz}(z)$ is the redshift distribution of the unknown sample, $\bar{b}_u(z)$ and $\bar{b}_r(z)$ are the evolution in bias of the unknown and reference samples, respectively, over the same scales, and $\bar{\omega}_{DM}(z)$ is the equivalent evolution in the integrated dark matter correlation function.

\subsection{The bias evolution of the unknown sample}
\label{sec:clustz_bias_eq}

In order to compute a redshift distribution, we need an estimate of $\bar{b}_u(z)$, $\bar{b}_r(z)$, and $\bar{\omega}_{DM}(z)$. Assuming linear biasing, the integrated autocorrelations of the unknown and reference samples as a function of redshift can be written as $\bar{\omega}_{uu}(z) = \bar{b}_{u}^{2}(z) \bar{\omega}_{DM}(z)$ and $\bar{\omega}_{rr}(z) = \bar{b}_{r}^{2}(z) \bar{\omega}_{DM}(z)$ respectively. We are able to measure both $\omega_{uu}(z)$ and $\omega_{rr}(z)$, so we can substitute these in to equation \ref{eq:nz1}, producing,

\begin{equation}
\frac{dN_{u}}{dz}(z) \propto \frac{\bar{\omega}_{ur}(z)}{\sqrt{\bar{\omega}_{uu}(z)\bar{\omega}_{rr}(z)}}
\label{eq:nz2}
\end{equation}

We can measure $\bar{\omega}_{ur}(z)$, the integrated crosscorrelation between the unknown sample and each bin in redshift of the reference sample, and also $\bar{\omega}_{rr}(z)$, the integrated autocorrelation of the reference sample over the same redshift bins and physical scale. We can also remove the constant of proportionality by normalising $\frac{dN_{u}}{dz}(z)$ to the number of galaxies in the unknown sample. The only parameter we cannot compute is $\bar{\omega}_{uu}(z)$, since we have no redshift information for the unknown sample. 

M13 show in their figure 1 that for a range of assumed bias evolutions of the unknown sample, if the redshift distribution is narrow, $\sigma_z < 0.2$, the effects of bias evolution on the recovered distribution are small, and therefore the distribution can be estimated as $\frac{dN_{u}}{dz}(z) \propto \bar{\omega}_{ur}(z)$. Some papers choose to assume this relation, e.g. M13, \cite{schm13}, or factor any deviation from this into their error budgets \cite{gatt18}.


\subsection{Estimating a bias correction}
\label{sec:clustz_bias_evo}

\textcolor{black}{In this work we apply the methodology described in the previous section to many different magnitude bins of galaxies and, as shown in section \ref{sec:test_clustz}, while distributions are generally narrow, they can often be wider than $\sigma_z = 0.2$. For this reason, we investigate the effect of different assumptions about the evolution of the bias of the unknown sample on our redshift distributions, and also in our final stellar mass, luminosity and completeness functions. This section describes our two different approaches: a correction based on the evolution of the clustering as measured in L-Galaxies, and analytic forms for the evolution of the bias.}

\subsubsection{Bias evolution from L-Galaxies}
\label{sec:clustz_bias_evo_LGal}
\textcolor{black}{One way to correct for the bias evolution of the unknown sample is to investigate how this evolves for different samples of galaxies in the LGalaxies semi-analytic model described in section \ref{sec:data_mock}. This approach has the advantage of being applicable to any sample of galaxies since we can look at the same bin of galaxies in our model and compute a correction. We will see, however, that while it should be correct for mock data, the derived correction may not hold for real data.}

\begin{figure*}
	\includegraphics[width=\linewidth]{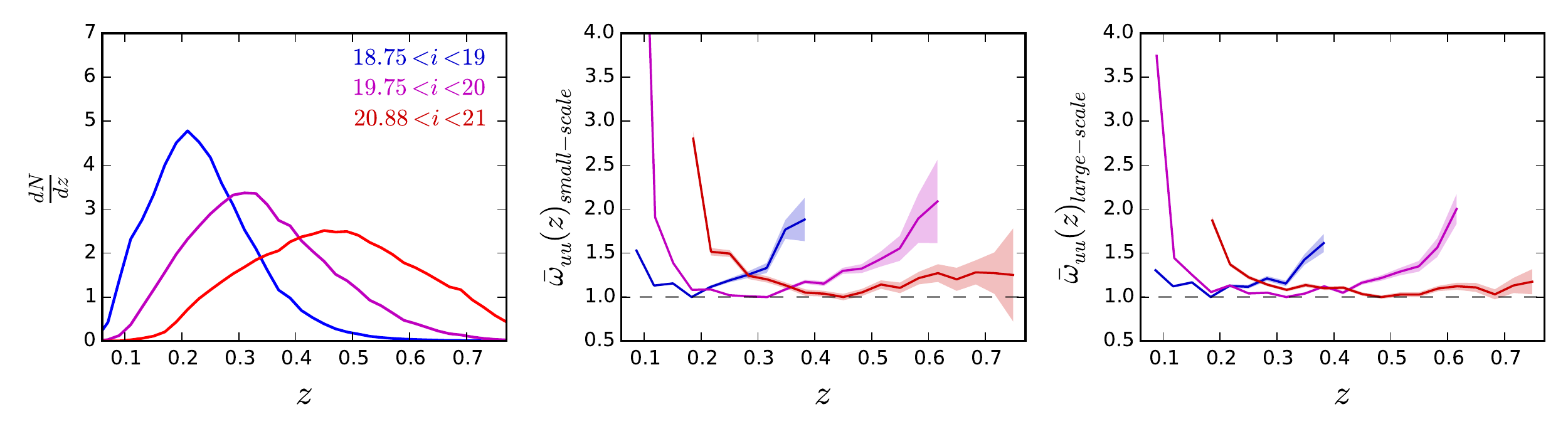}
    \caption{(left) The normalised redshift distributions of three different magnitude bins of LGalaxies data: $18.75 < i < 19$ (blue), $19.75 < i < 20$ (magenta), and $20.875 < i < 21$ (red). The middle and right hand side plots show the integrated angular autocorrelation function, $\bar{\omega}_{uu}(z) = \bar{b}_{u}^{2}(z) \bar{\omega}_{DM}(z)$, as a function of redshift for these three bins. The distributions are shown for small scales, 0.5 < $r_p$ < 1.5 Mpc (middle) and large scales, 5 < $r_p$ < 15 Mpc, (right). Since the clustering redshifts method only depends on the evolution of $\bar{\omega}_{uu}(z)$, not the overall amplitude, integrated correlation functions are normalised such that the minimum value is 1. Error bars are computed from correlation function errors.}
    \label{fig:bias_evo}
\end{figure*}

We first measure the integrated autocorrelation, $\bar{\omega}_{uu}(z)$ as a function of redshift in L-Galaxies data. We measure $\bar{\omega}_{uu}(z)$ for 10 bins of $i$-band magnitude of width $\Delta i = 0.25$ between $17 < i < 20$ and $\Delta i = 0.125$ between $20 < i < 21$ (since we have significantly more galaxies at fainter magnitudes). For each magnitude bin, we measure $\bar{\omega}_{uu}(z)$ in redshift bins of width $\Delta z = 0.1/3$, as this is the binning we will apply to test our data in section \ref{sec:test_clustz}. Measuring $\bar{\omega}_{uu}(z)$ in galaxies with different magnitudes accounts for any evolution of the bias correction as a function of luminosity. Figure \ref{fig:bias_evo} shows $\bar{\omega}_{uu}(z)$ computed in three different magnitude bins. Two separate panels show the integrated correlation function amplitude over either small scales (0.5<r$_p$<1.5 Mpc, middle panel) or large scales (5<r$_p$<15 Mpc, right panel). 


Looking at both small scale and large scale clustering, in all bins of magnitude, there is a significant increase in the clustering amplitude towards lower redshifts. This behavior is particularly noticeable in the two faintest magnitude bins. There is also, in all magnitude bins, an increase in clustering amplitude towards higher redshifts, although in general this evolution is smaller at larger scales. 


\textcolor{black}{Some of the evolution in integrated clustering amplitude seen in Figure \ref{fig:bias_evo} is expected from the fact that we have magnitude-limited samples. For a given magnitude bin, galaxies at high redshift are, on average, intrinsically more luminous and therefore more massive and more strongly biased. This will cause an increase in clustering amplitude towards high redshifts, as seen in Figure \ref{fig:bias_evo} in all magnitude bins.}

\textcolor{black}{This does not, however, explain the increase towards low redshift, where galaxies are of lower luminosity and stellar mass, hence not strongly biased. We investigate a number of reasons for this amplitude increase, including the fact that we are measuring the evolution of $\bar{\omega}_{uu}(z)$, which captures the evolution of both the bias of the unknown sample and of the dark matter power spectrum (i.e. $\bar{\omega}_{DM}(z)$), the latter of which increases in amplitude towards low redshift, We also investigate the effect of a changing satellite fraction with stellar mass on the clustering strength. We find, however, that these effects are not significant, and the main reason for this increased amplitude is due to the intrinsic clustering properties of the model. \cite{vand16} present an exploration of how the clustering of galaxies can aid the constraint of semi-analytic models. In their Fig. 5 they show that without explicitly using clustering as a model constraint, several flavours of L-Galaxies models fail to reproduce the clustering of low-mass galaxies ($M_{\star} \lesssim 10^{9.5} M_{\odot}$), even if the clustering of high mass galaxies matches the SDSS-measured correlation functions very well. The tendency for L-Galaxies to over-predict the clustering amplitude of low-mass galaxies (the model that we use here is not calibrated using clustering) will produce the behaviour seen at low redshift in Figure \ref{fig:bias_evo}.}

When applying a bias correction from L-Galaxies in future sections, we use the measured evolution of $\bar{\omega}_{uu}(z)$ in $SDSS_{LGalaxies}$ as an estimate of the bias evolution of the unknown sample, following equation \ref{eq:nz2}. This correction is computed in the same magnitude bin and over the same physical scales as the cross-correlation is measured. 

\subsubsection{Analytic bias evolution}
\label{sec:clustz_bias_evo_2}


\textcolor{black}{We will also consider analytic forms of the bias, when investigating the effect of the bias evolution of the unknown sample. 
\cite{rahm15} compute bias corrections fit to SDSS main spectroscopic sample clustering: one that evolves as $db/dz = 1$ and one that evolves as $db/dz = 2$. For our study we choose the more extreme evolution, $b_1(z) = 0.7 + 2z$, in an effort to bound the effect of this uncertainty. We also investigate a correction from \cite{rahm16b}, which takes the form $b_2(z) = 1$ for $z<0.1$, and $b_2(z) = 0.9 + z$ for $z \geq 0.1$. }

\subsubsection{Quantifying the effect of the bias evolution of the unknown sample}

\textcolor{black}{In Sections \ref{sec:testmeth2} and \ref{sec:mlfunc} we will compute redshift distributions, luminosity functions, stellar mass functions and completeness functions with different assumptions about the bias evolution of the unknown sample. When testing our methodology in simulated data, we will apply the bias evolution derived in section \ref{sec:clustz_bias_evo_LGal}, and quantify the effect of not correcting for this evolution at all. However, due to the shortcomings of L-Galaxies in describing the evolution of the clustering of low-mass galaxies, when analysing real data we will also consider the analytic forms described in Section~\ref{sec:clustz_bias_evo_2}. Ultimately, we will use the these models to quantify the likely effect of the bias evolution of the unknown sample in our final measurements, and add this systematic error to our estimates of the statistical error.}


\subsection{Clustering measurements}

\textcolor{black}{Evaluating equation \ref{eq:nz2} requires us to make choices regarding correlation function estimators, cosmological parameters, and scales over which to integrate correlation function amplitudes. Although large-scale clustering is less dependent on assumptions about the bias evolution of the unknown sample, the recovered $\phi(z)$ is significantly noisier, mostly due to the angular cross-correlation signal being diluted by foreground/background galaxies. Furthermore, this signal is more susceptible to spurious correlations due to large-scale structure (e.g., chance alignment of structure at different redshifts). We discuss this issue in appendix \ref{sec:append_1}. Because of this effect, when applying our method to real data we choose to measure cross-correlations over the scales 1.5 to 5 Mpc, as 1.5 Mpc is the smallest scale we can measure while being safely above the SDSS fiber collision radius of $62$ arcseconds. We use the \cite{land93} estimator in all our correlation function measurements. Furthermore, where cosmology is needed we assume \cite{plan14} best fit cosmological parameters.}

\subsection{Computing Masses and Luminosities}
\label{sec:clustz_mass_lums}

Before recovering redshift distributions, we seek to bin galaxies in small bins of colour and magnitude (i.e. binned in three dimensions by $i$, $r-i$ and $g-r$). This binning is useful because firstly, since galaxy colours are strongly correlated with redshift, it limits the width of redshift distributions for each bin, which will in turn reduce the importance of the bias evolution correction \citep{newm08, mena13}. Secondly, since we intend to compute masses and luminosities, we require both colour and redshift information, so binning by colour is important.

After recovering the redshift distributions of galaxies in all these colour bins, for each bin, we know the value of $i$, $r-i$ and $g-r$, along with the number of galaxies at each redshift. At each redshift, we therefore have a measure of the rest-frame spectral energy distribution (SED) of galaxies in this bin. We can compute parameters from this SED (e.g. mass, luminosity) and allocate these to photometric galaxies in the correct quantities.

To compute masses and luminosities for our bins of colour, we choose to use semi-analytic models. After recovering redshift distributions of all bins of colour of our photometric survey, we compute the distribution of mass or luminosity within the same colour-redshift bin of the SAM, and apply this probability density function (PDF) to the real data (i.e. multiply this PDF by the number of galaxies in the same bin of the real data). After recovering masses and luminosities at each redshift, for all bins of colour, these distributions are summed to produce mass and luminosity functions. 

Although SAMs do not predict the correct number density of galaxies of given colours, for a given set of colours and redshift, the type of galaxy (i.e. star formation history (SFH), mass, luminosity) should be representative of those in the real universe. This method allows us to account for photometric errors by adding these to our SAM, and to produce a PDF of mass and luminosity rather than just a best fit, ensuring that the correct distribution of mass is allocated in each bin. This technique is, in essence, similar to \cite{paci12}, where a library of physically motivated SFHs is computed from SAMs, and then used to fit individual galaxy SEDs.

\section{Testing the methods}

\label{sec:testmeth2}

Before computing real mass and luminosity functions, we test our method using real and mock data. We create mock reference samples and photometric surveys as described section \ref{sec:data_mock}. This allows us to select galaxies from our photometric survey as function of magnitude and colour, and recover redshift distributions by crosscorrelating with the reference sample. We can then compare the recovered redshift distributions to the true distribution.

\subsection{Clustering Redshifts on Mock Data}
\label{sec:test_clustz}

In order to test the clustering redshifts method, we bin mock photometric survey, $SDSS_{LGalaxies}$ by $i$-band magnitude in bins of width $\Delta i = 0.25$ between $17 < i < 20$ and $\Delta i = 0.125$ between $20 < i < 21$ where the large number of galaxies allows us to bin more finely. Within each of these magnitude bins, we then bin by $r-i$, and then by $g-r$. We choose a number of bins such that each contains $>100,000$ galaxies, as we found this to be roughly the minimum number of galaxies required to recover a noise-free $n(z)$. At fainter magnitudes, the size of bins is comparable to the photometric error in the SDSS, so smaller bins would not provide significantly more information as galaxies are already scattered between bins. Binning by $i$, $r-i$ and $g-r$ produces 492 bins.

\begin{figure*}
	\includegraphics[width=285pt]{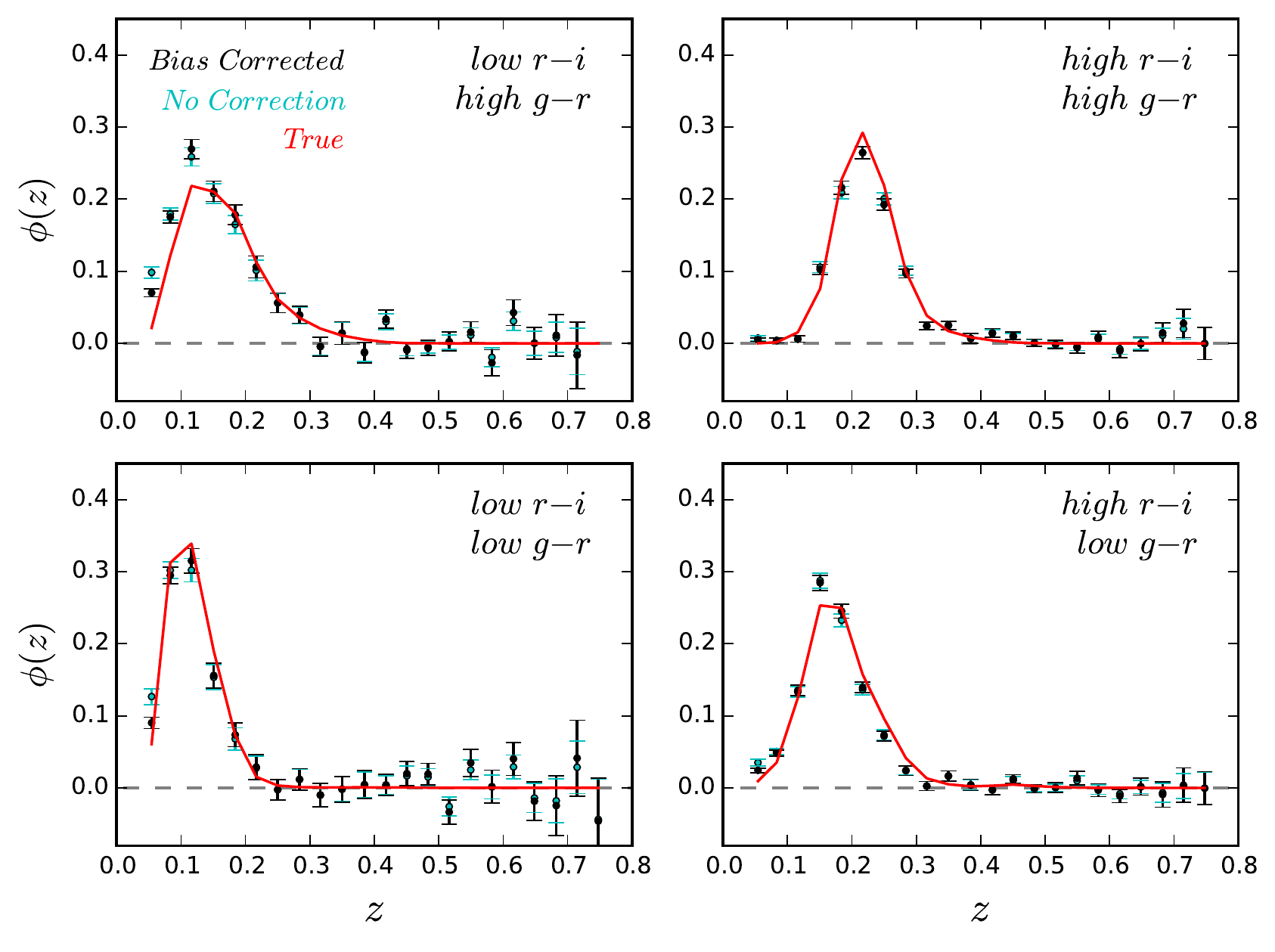}
    \caption{The recovered redshift distributions of different bins of colour of SDSS$_{LGalaxies}$ data, with both no bias correction (cyan), and the bias correction computed in section \ref{sec:clustz_bias_evo} (black). The true distribution is given by the red line. We choose galaxies from a bright magnitude bin ($18 < i < 18.25$), and show four colour bins covering the extent of the colour space.}
    \label{fig:phi_z_tests_2}
\end{figure*}

\begin{figure*}
	\includegraphics[width=465pt]{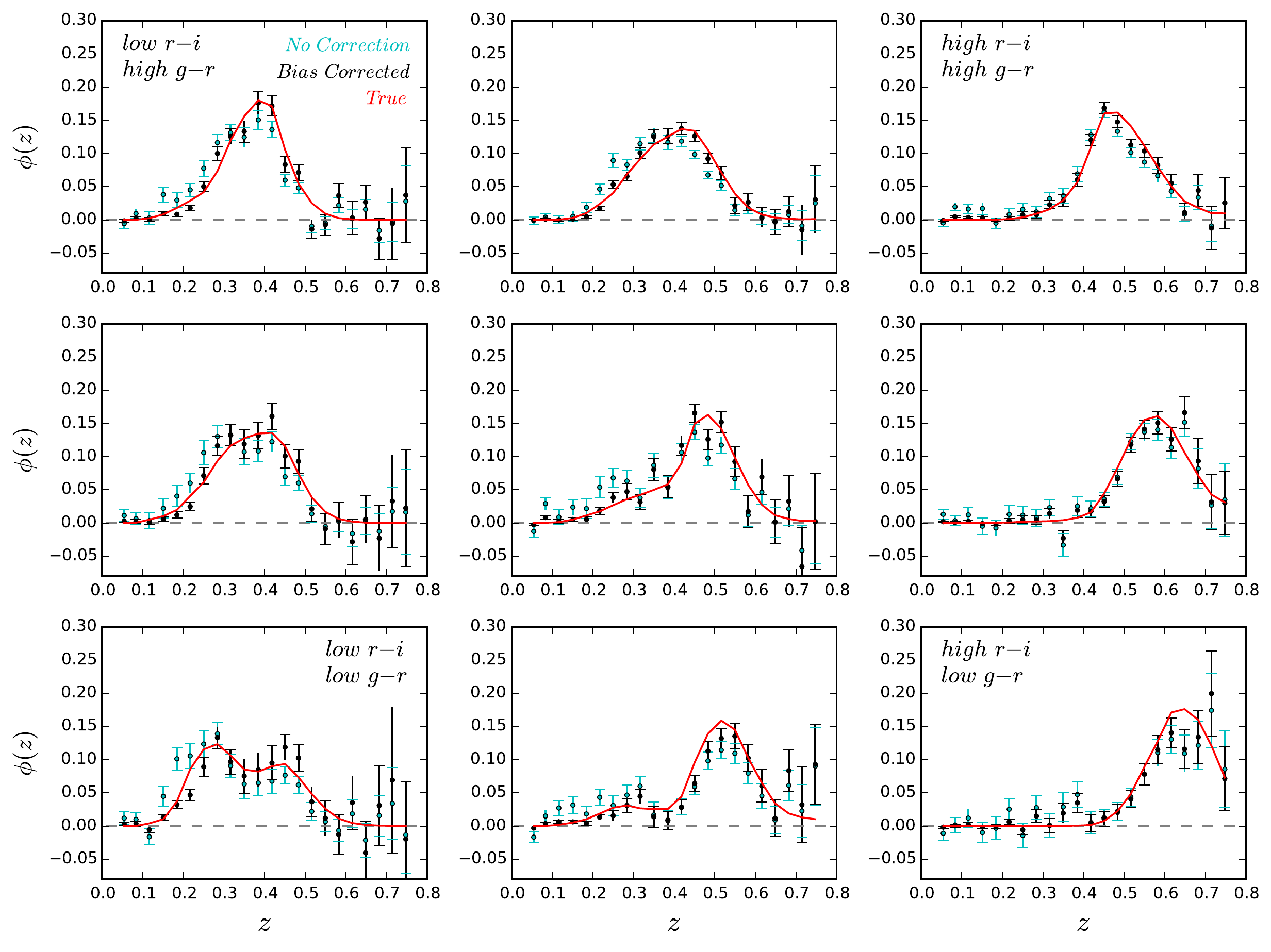}
    \caption{The recovered redshift distributions of different bins of colour of SDSS$_{LGalaxies}$ data, with no bias correction (cyan), and the bias correction computed in section \ref{sec:clustz_bias_evo} (black). The true distribution is given by the red line. Here we choose galaxies from the faintest magnitude bin ($20.875 < i < 21$), containing 7x7 bins in $r-i$ and $g-r$. A selection of bins is presented throughout the colour space (bins 2, 4 and 6 in both dimensions). When computing redshift distributions, the crosscorrelation is integrated over small scales (0.5 < $r_p$ < 5 Mpc).}
    \label{fig:phi_z_tests}
\end{figure*}

\begin{figure}
\includegraphics[width=\linewidth]{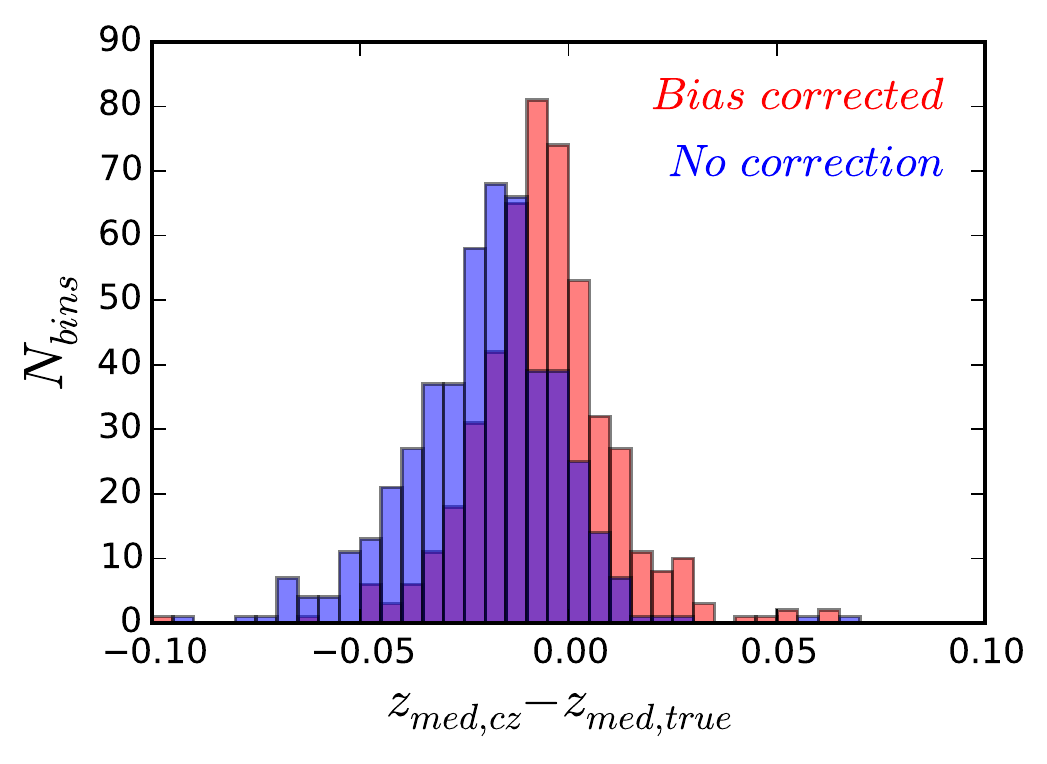}
    \caption{Testing the recovery of redshift distributions in L-Galaxies. The distribution of errors in the median redshift, $z_{med,true} - z_{med,cz}$, for all bins of colour and magnitude, both with and without the bias correction. }
    \label{fig:median_offsets}
\end{figure}

We then recover the redshift distributions of all bins of $SDSS_{LGalaxies}$ by crosscorrelating with a reference sample, $BOSS_{LGalaxies}$, as described in sections \ref{sec:clustz_meth} and \ref{sec:clustz_bias_eq}. We correct for bias evolution using the computed evolution in LGalaxies as in section \ref{sec:clustz_bias_evo_LGal}. Correlation functions errors are computed using a jacknife method, which in turn is used to compute errors on the final $dN/dz$ following equations \ref{eq:int_cross} and \ref{eq:nz2}. Figures \ref{fig:phi_z_tests_2} and \ref{fig:phi_z_tests} show the recovered and true redshift distributions of a selection of these colour bins, in bright and faint magnitude bins, respectively.

Figure \ref{fig:phi_z_tests_2} shows four bins of $r-i$ and $g-r$ covering the extent of the colour space. It can be seen that the redshift distribution, $\phi(z)s$, is recovered well for a range of different values of $g-r$ and $r-i$. Adding a bias correction does not significantly affect the recovered distribution, likely because distributions are narrow, and because the correction, computed in section \ref{sec:clustz_bias_evo_LGal}, is fairly small at bright magnitudes. Examining the faintest magnitude bin in figure \ref{fig:phi_z_tests}, redshift distributions are again recovered well for a range of different values of $g-r$ and $r-i$, however the bias evolution correction becomes more important. This effect appears to be particularly true for wider distributions, where the bias is likely changing between low and high redshift following figure \ref{fig:bias_evo}.

If using a photometric survey with smaller photometric error, for example DECaLS or DES, redshift distributions for a given colour bin would be much narrower since galaxies will be less scattered between neighboring bins. The correction therefore becomes less significant, particularly at faint magnitudes, where SDSS errors are large. Errors are visibly larger at higher redshift (z > 0.65), where the number density of objects in the reference sample is low, which can sometimes cause an error in normalisation. This effect should average out over many bins, however, and will be less of a problem when using real data since the true BOSS sample has a larger area, and there are additional eBOSS galaxies and quasars above this redshift.

As a further check of our bias correction, we compute the true median redshift, $z_{med,true}$, for all colour bins, along with the median redshift using clustering redshifts, $z_{med,cz}$, both with and without a bias correction. We compute the error in the median redshift, $z_{med,true} - z_{med,cz}$, for all bins, and present the distribution of errors in figure \ref{fig:median_offsets}.

Without a bias correction, median redshifts are almost always slightly below the true value. The bias correction shifts the median to higher redshifts, although there remains a similar amount of scatter around the correct value. These errors in the median redshift are fairly small however, relative to the size of our redshift bins ($\Delta z = 0.033$). The scatter is partly due to to noise in the recovered redshift distribution, but also may arise because we compute a bias correction for an entire magnitude bin, and this approach may not necessarily describe the bias evolution of all bins of colour within this.

\subsection{Clustering Redshifts on Real Data}
\label{sec:test_clustz_real_data}

\begin{figure*}
	\includegraphics[width=\linewidth]{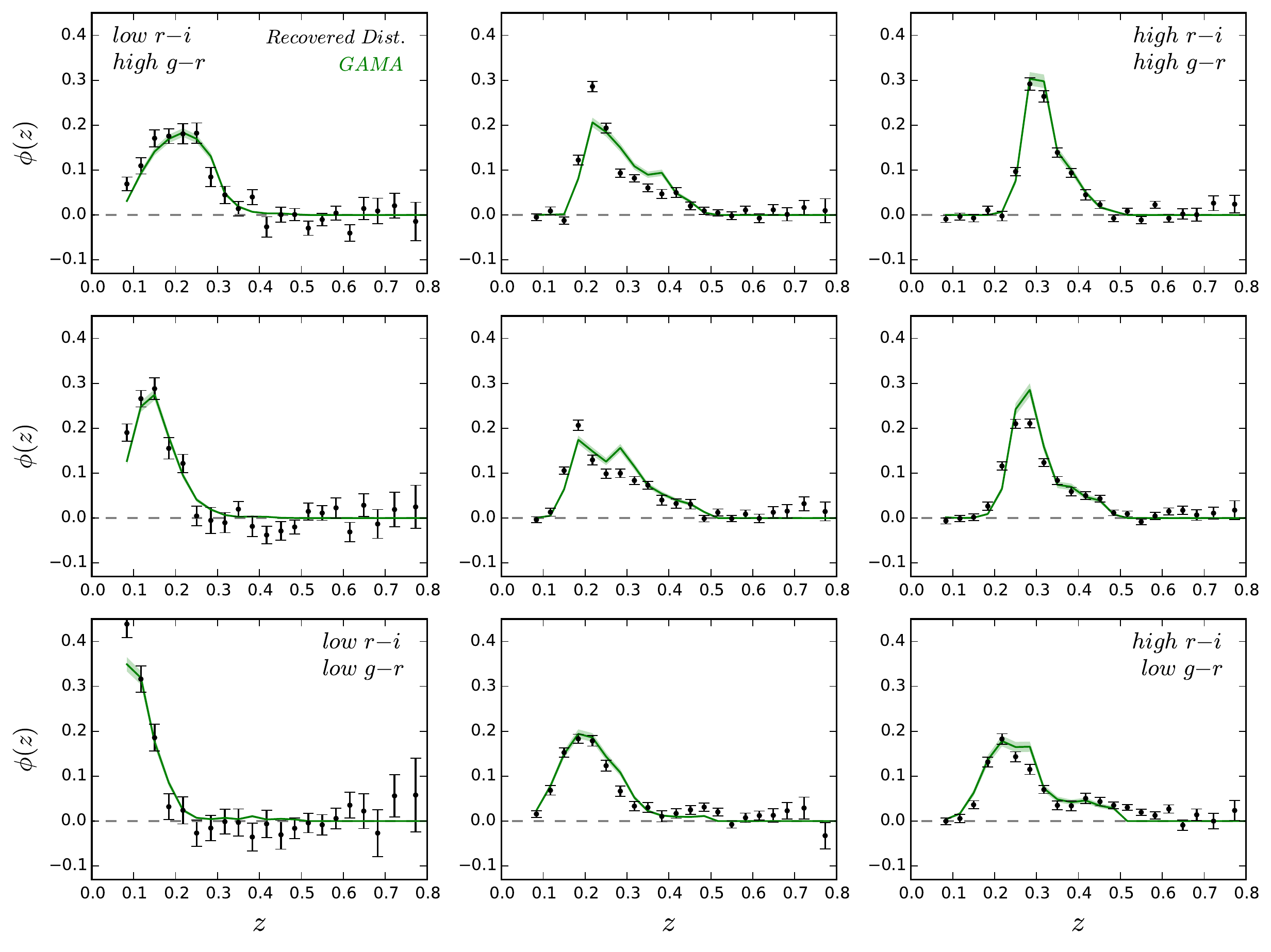}
    \caption{The recovered redshift distributions of different bins of colour of real data both with (black points), and without (cyan points) a correction. The spectroscopic redshift distribution of GAMA galaxies is indicated in the same colour bin (green line). We choose the magnitude bin $(19 < i < 19.25)$, and only show bins with small values of $r-i$ in order to avoid the $r<19.8$ magnitude cut in GAMA.}
    \label{fig:phi_z_tests_real_data}
\end{figure*}

\textcolor{black}{In this section we test our method using a highly complete set of spectroscopic redshifts, using data from the GAlaxy and Mass Assembly spectroscopic redshift GAMA \citep{driv09}. GAMA is a spectroscopic survey magnitude limited to $r < 19.8$, targeted over $\sim$286 deg$^{2}$ of sky, its primary objective being to study structure on scales of 1 kpc to 1 Mpc. Below $r < 19.8$, GAMA is highly complete ($> 95\%$), although completeness drops for fainter magnitudes. Therefore, for $r \lesssim 19.8$, GAMA redshift distributions should be roughly comparable to the SDSS.}

\textcolor{black}{We use a photometric survey defined from SDSS data, again cut to $17 < i < 21$, described in section \ref{sec:data_phot}. As in section \ref{sec:test_clustz}, we split the sample into bins of $i$-band magnitude of width $\Delta i = 0.25$ between $17 < i < 20$ and $\Delta i = 0.125$ between $20 < i < 21$, and then bin by $r-i$ and $g-r$ within each of these such that each bin contains $>100,000$ galaxies. We cross-correlate each of these bins with our reference sample described in section \ref{sec:data_ref}, consisting of BOSS and eBOSS LRGs and quasars, in order to recover redshift distributions. An example of some recovered distributions is presented in Figure \ref{fig:phi_z_tests_real_data}, alongside the redshift distributions measured from the GAMA survey \citep{bald18} in the same bins of magnitude and colour. We choose an intermediate magnitude bin, $19 < i < 19.25$, in order that we have galaxies over a range of redshifts. In order to lessen the effect of the $r$-band magnitude cut in GAMA, we only show the bluest bins such that bins have $r \lesssim 19.8$, where incompleteness is not significant. Clustering redshifts recovery is shown without any bias correction, since in our tests, the correction is not significant in these bins.}


\textcolor{black}{Recoveries of SDSS redshift distributions generally match the corresponding GAMA colour bin well. Some small differences are visible; however, this was also true for the simulated data in figure \ref{fig:phi_z_tests}, for which the mass function is recovered well. If we take only bins below i < 19.25, we can use GAMA to compute the error in the median redshift of each colour bin, $z_{med,GAMA} - z_{med,cz}$, as in section \ref{sec:test_clustz}. After computing this for all bins, the average error is $\delta z_{med} = -0.01$, indicating no significant offset with the spectroscopic redshift distribution.}

\subsection{Mass \& Luminosity Functions of Mock Data}
\label{sec:test_mf_lf}

\begin{figure*}
\includegraphics[width=\linewidth]{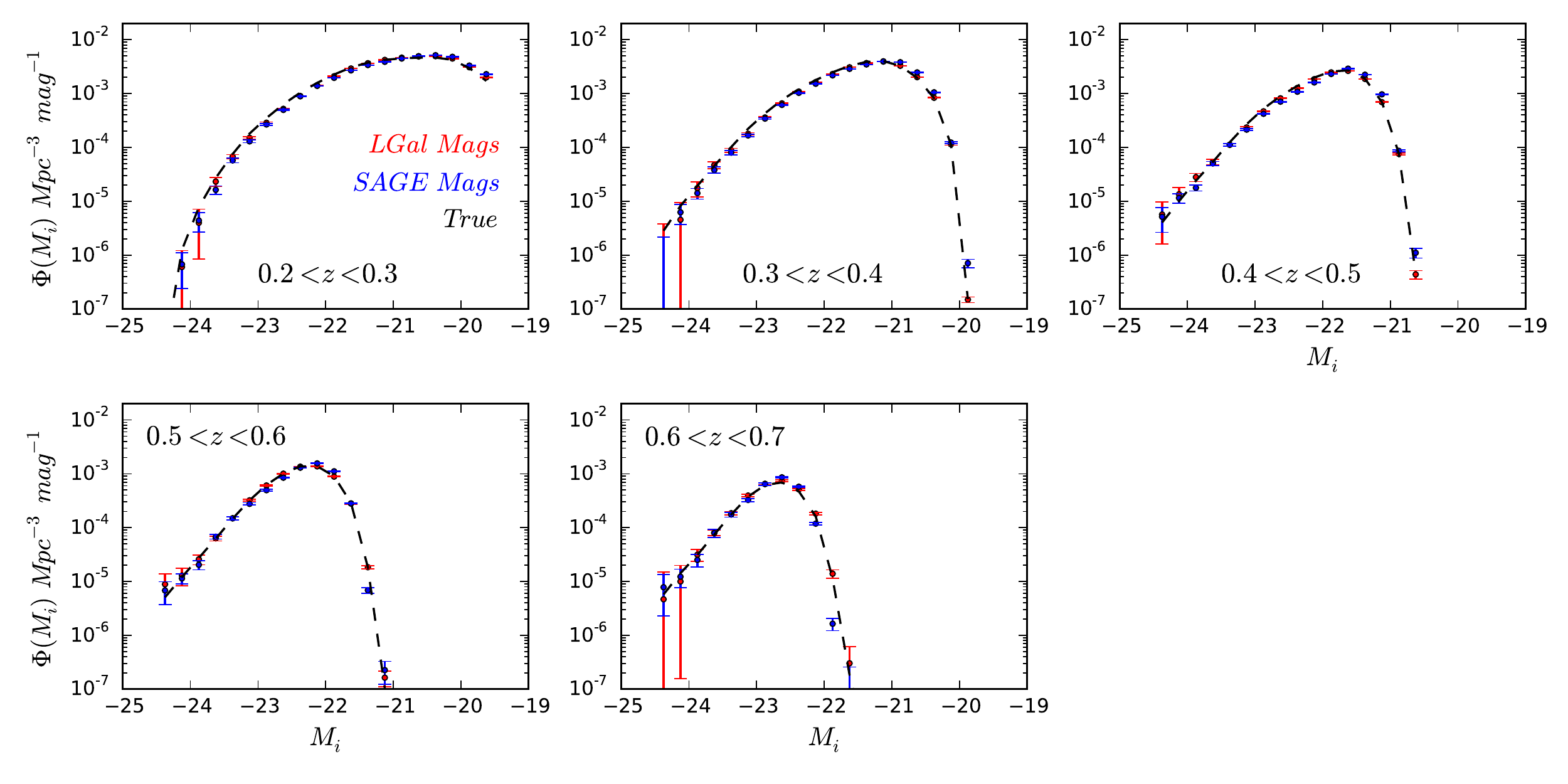}
    \caption{The recovered i-band k-corrected luminosity function of SDSS$_{LGalaxies}$ data for a number of different redshift bins between 0.2 and 0.7. Absolute magnitudes are recovered using colour-luminosity relations from both LGalaxies (red) and SAGE (blue) as described in section \ref{sec:test_mf_lf}, and the true luminosity function is shown as the black dotted line. The fall-off in the luminosity functions towards faint magnitudes is due to the i<21 cut in our sample.}
    \label{fig:lum_recovery}
\end{figure*}

\begin{figure*}
\includegraphics[width=\linewidth]{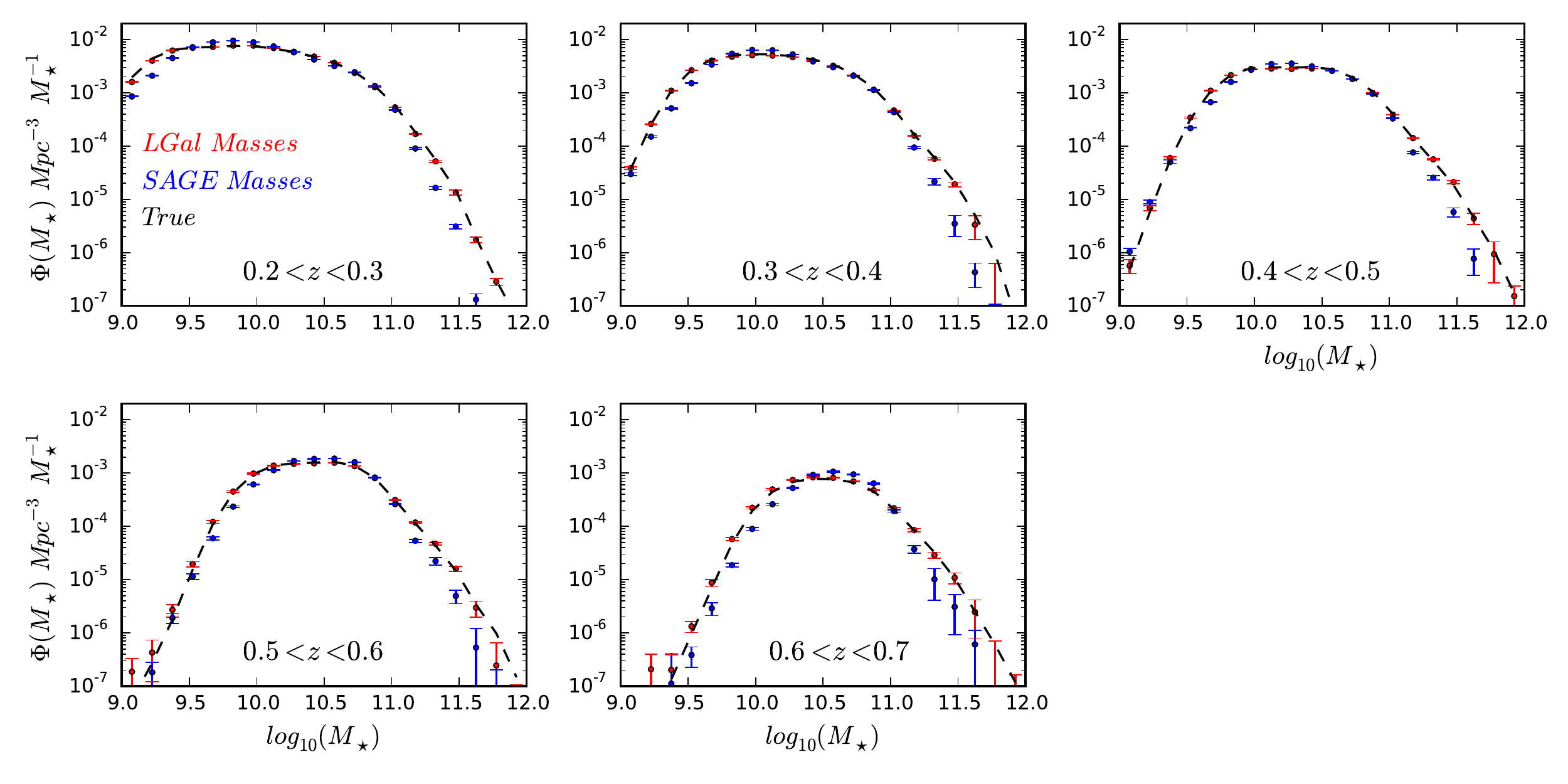}
    \caption{The recovered stellar mass function of SDSS$_{LGalaxies}$ data for a number of different redshift bins between 0.2 and 0.7. Masses are recovered using colour-mass relations from both LGalaxies (red) and SAGE (blue) as described in section \ref{sec:test_mf_lf}, and the true stellar mass function is shown as the black dotted line. The fall-off in the mass functions towards lower masses is due to the i<21 cut in our sample.}
    \label{fig:mass_recovery}
\end{figure*}

We now test our method of computing masses and luminosities, described in section \ref{sec:clustz_mass_lums}, on mock data. As in section \ref{sec:test_clustz}, we bin our mock photometric survey, $SDSS_{LGalaxies}$, in to bins of colour and magnitude, recovering redshift distribution in each. We then take each of these bins at a given redshift and allocate masses and luminosities by looking in both LGalaxies (the same model, but with different photometric noise applied), and smaller lightcones from SAGE (a different model, also with photometric noise applied). This approach tests how much the choice of model affects the estimated stellar masses and luminosities. After summing the mass and luminosity distributions for all bins of colour and redshift, we produce mass and luminosity functions between $17 < i < 21$. Errors are computed using the error in $\phi(z)$ from the clustering redshifts method. The recovered luminosity and mass function are shown in figures \ref{fig:lum_recovery} and \ref{fig:mass_recovery}.

Figure \ref{fig:lum_recovery} displays the recovery of luminosity functions of our $SDSS_{LGalaxies}$ survey, in different redshift bins. Since the luminosities allocated to our galaxies are in the rest-frame, the recovered luminosity functions are by definition k-corrected. We use both LGalaxies and SAGE to compute luminosities. The true luminosity function is recovered well at all redshifts, independent of whether LGalaxies or SAGE is used to compute an absolute magnitude. This result makes sense, since an absolute magnitude depends only on the redshift, cosmological model, and galaxy SED. Since we have accurate recovered redshifts and $i$, $r-i$ and $g-r$, we have effectively a rest frame SED, so the computed magnitude from this should not be particularly dependent on the SAM chosen.

Figure \ref{fig:mass_recovery} shows mass functions, again recovered at different redshifts for the two different models. Using LGalaxies to recover masses works very well (i.e., the same model to convert colours and redshifts to masses), with the recovered mass functions almost exactly matching the true values at all redshifts and masses. Examining the SAGE results, at $M_{\star} < 10^{11.25} M_{\odot}$, mass functions are recovered well; however, above these masses, the number of high mass galaxies is under-predicted.

\begin{figure*}
\includegraphics[width=\linewidth]{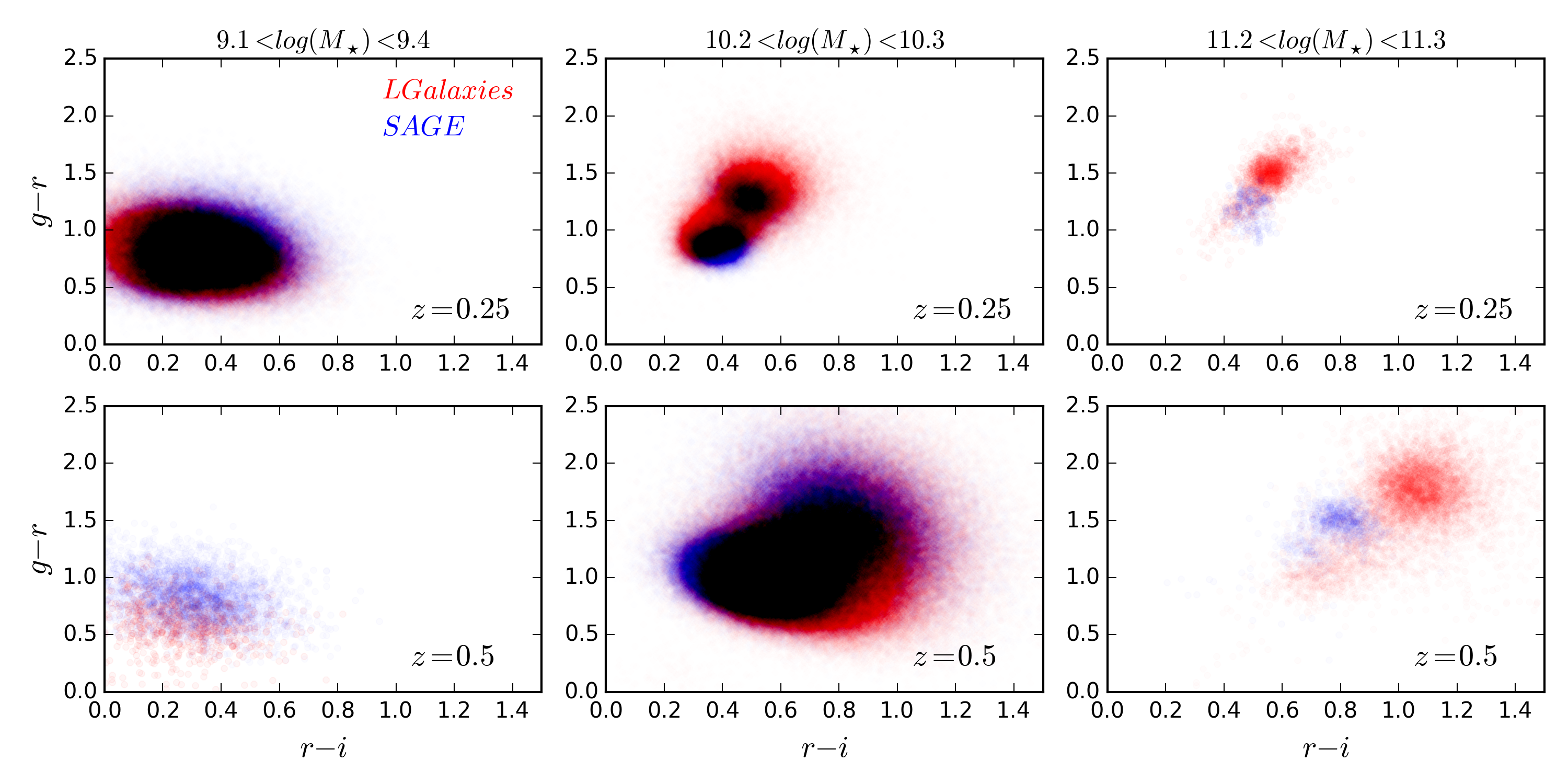}
    \caption{The $r-i$ and $g-r$ colours of galaxies in LGalaxies (red) and SAGE (blue). This distribution is shown for two different redshifts, 0.25 and 0.5, and for three bins of mass centered around 9.25, 10.25 and 11.25 $\log(M_{\odot})$.}
    \label{fig:lgal_sage_colours}
\end{figure*}

\begin{figure}
\includegraphics[width=\linewidth]{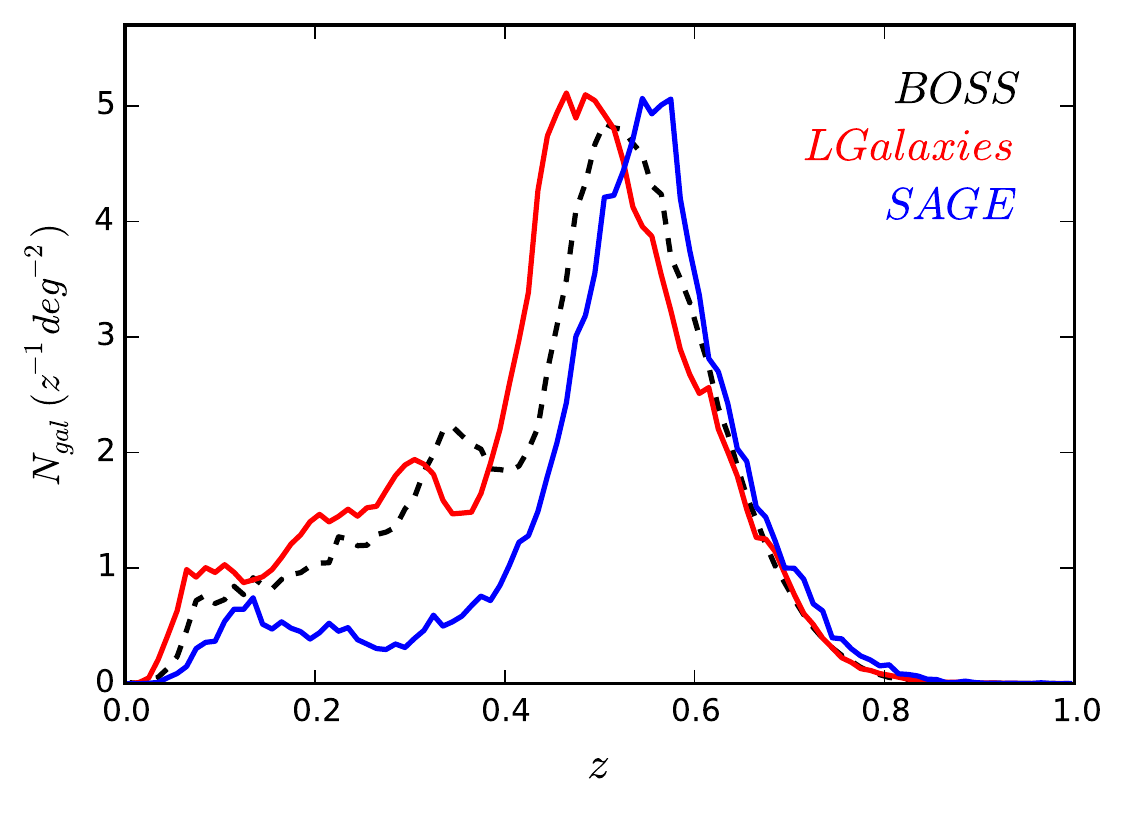}
    \caption{The redshift distribution (number of galaxies deg$^{-2}$ z$^{-1}$) for the BOSS survey (black dotted line) compared with redshift distributions of LGalaxies (red) and SAGE (blue) with the BOSS colour cuts.}
    \label{fig:lgal_sage_boss}
\end{figure}

In order to understand this difference, we compare the distribution of colours in both models as a function of mass in figure \ref{fig:lgal_sage_colours}. In the two lowest mass bins ($M_{\star}$ = $10^{9.25} M_{\odot}$ and $10^{10.25} M_{\odot}$), both SAGE and LGalaxies cover roughly the same colour space at both redshifts ($z=0.25$ and $0.5$). This result implies that colours of low mass galaxies ($M_{\star} \lesssim 10^{11} M_{\odot}$) are fairly independent of the semi-analytic model chosen, and explains why the mass function is recovered well at lower masses. In the high mass bin ($M_{\star} = 10^{11.25} M_{\odot}$), colours are visibly different in the two models. This behavior implies that high mass galaxies likely have different formation processes in the two models, and explains why mass functions are not recovered as well.

Since we do not know exactly which model best describes the real universe at high masses, we investigate how well both can reproduce the BOSS survey (containing large numbers of massive galaxies). We apply the colour cuts of BOSS to both samples as described in section \ref{sec:data_mock}, and compare the redshift distributions of these samples and the real BOSS survey in figure \ref{fig:lgal_sage_boss}. It can be seen that LGalaxies reproduces both samples within the BOSS survey: the LOWZ sample at $0<z<0.4$, and CMASS sample at $0.4<z<0.8$. These are recovered with broadly the same number density, and although there is a slight offset in the peak of the sample, the overall shape of both samples is recovered well. SAGE manages to select some galaxies with a distribution similar to CMASS; however, at low redshift most galaxies are missing, and the overall shape is significantly different.

For this reason we trust that the colours of galaxies are significantly closer to those of the real universe in LGalaxies. We therefore opt to use LGalaxies when computing masses of real data in section \ref{sec:mlfunc}.

\subsection{The effect of a bias correction on stellar mass functions from simulated data}
\label{sec:test_bias_corr}
\begin{figure}
\includegraphics[width=\linewidth]{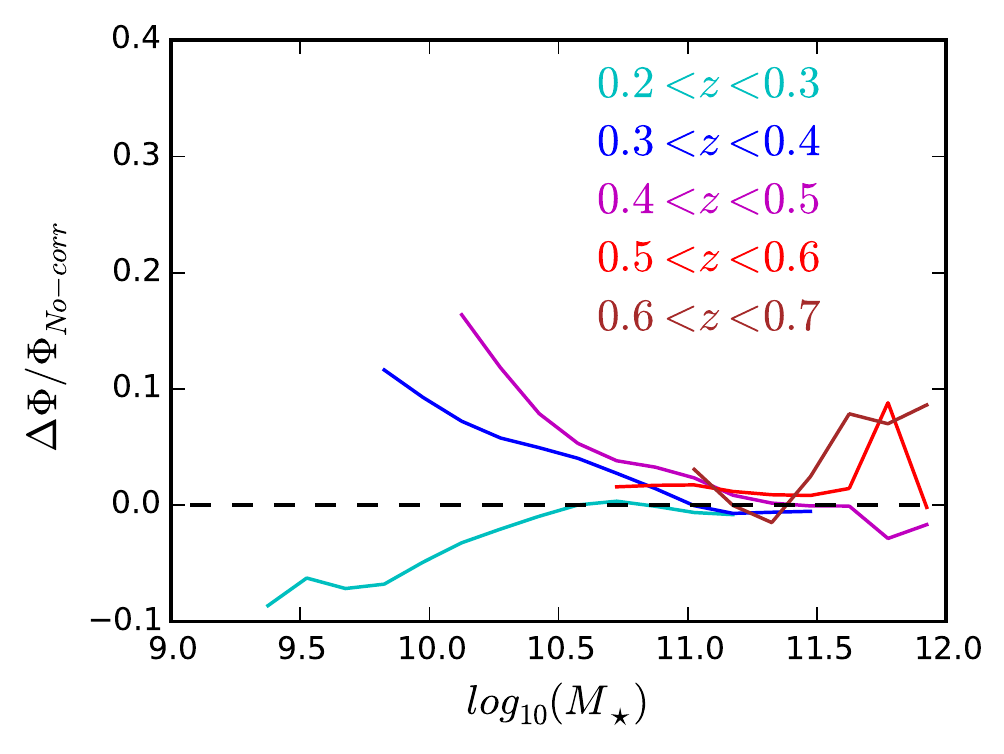}
    \caption{The fractional difference in the mass function between using an unknown sample bias correction from SAMs and no-correction, plotted for masses where mass functions are more than 95\% complete as in section \ref{sec:mfunc}}
    \label{fig:mock_mf_corr}
\end{figure}

\textcolor{black}{We now test the importance of the bias correction on the recovered mock mass functions. To do this we compute mass functions with and without the bias correction detailed in section \ref{sec:clustz_bias_evo_LGal}. We then compute the ratio of these mass functions $\frac{\Phi(M_{\star})}{\Phi(M_{\star})_{nc}} - 1$, where $\Phi(M_{\star})_{nc}$ represents the mass function without bias correction. This quantity shows the fractional change in mass function when using a bias correction compared with no correction. We show this quantity for different redshift bins in figure \ref{fig:mock_mf_corr}.}

\textcolor{black}{The effect of the bias correction is more pronounced at lower masses $M_{\star} < 10^{10.5} M_{\odot}$; at larger masses the change is only of the order of a few percent. We will see later, in Section~\ref{sec:mfunc_b_corr}, that this matches well with similar tests in the data, and that these uncertainties are comparable to our statistical error.}



\section{Mass and Luminosity Functions of SDSS Data}
\label{sec:mlfunc}

\begin{figure*}
\includegraphics[width=\linewidth]{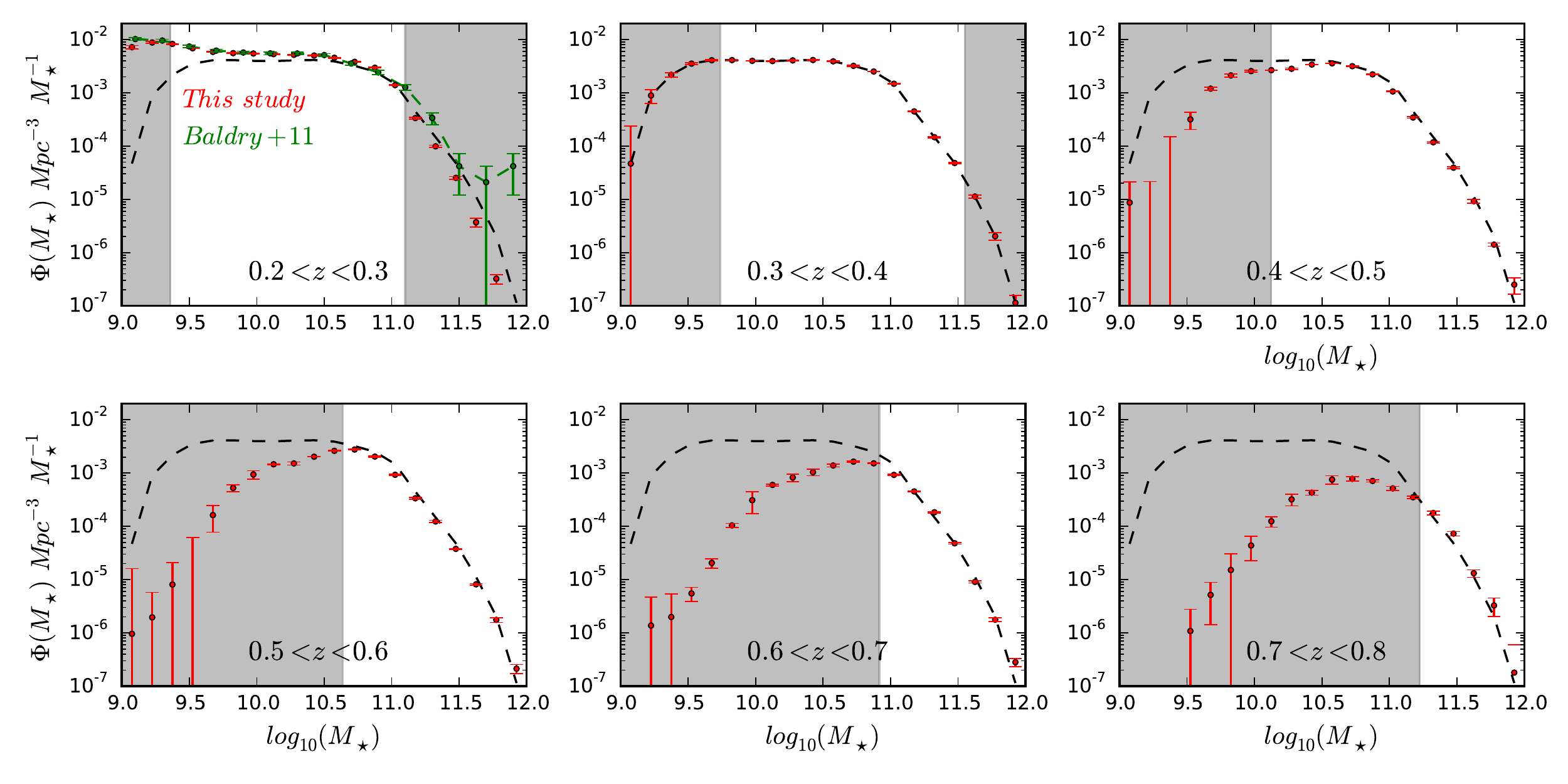}
    \caption{Recovered mass functions for real SDSS data in a number of different redshift bins (red). The green points in the redshift bin 0.2 < z < 0.3 are the GAMA mass function (z < 0.06) \citep{bald12}. For reference, the mass function computed using our method in the 0.3 < z < 0.4 bin is shown in all bins as the black dashed line. Regions where our mass functions are more than 95\% incomplete (due to the photometric sample magnitude cut) are shown in grey.}
    \label{fig:sdss_mf}
\end{figure*}

\begin{figure*}
\includegraphics[width=\linewidth]{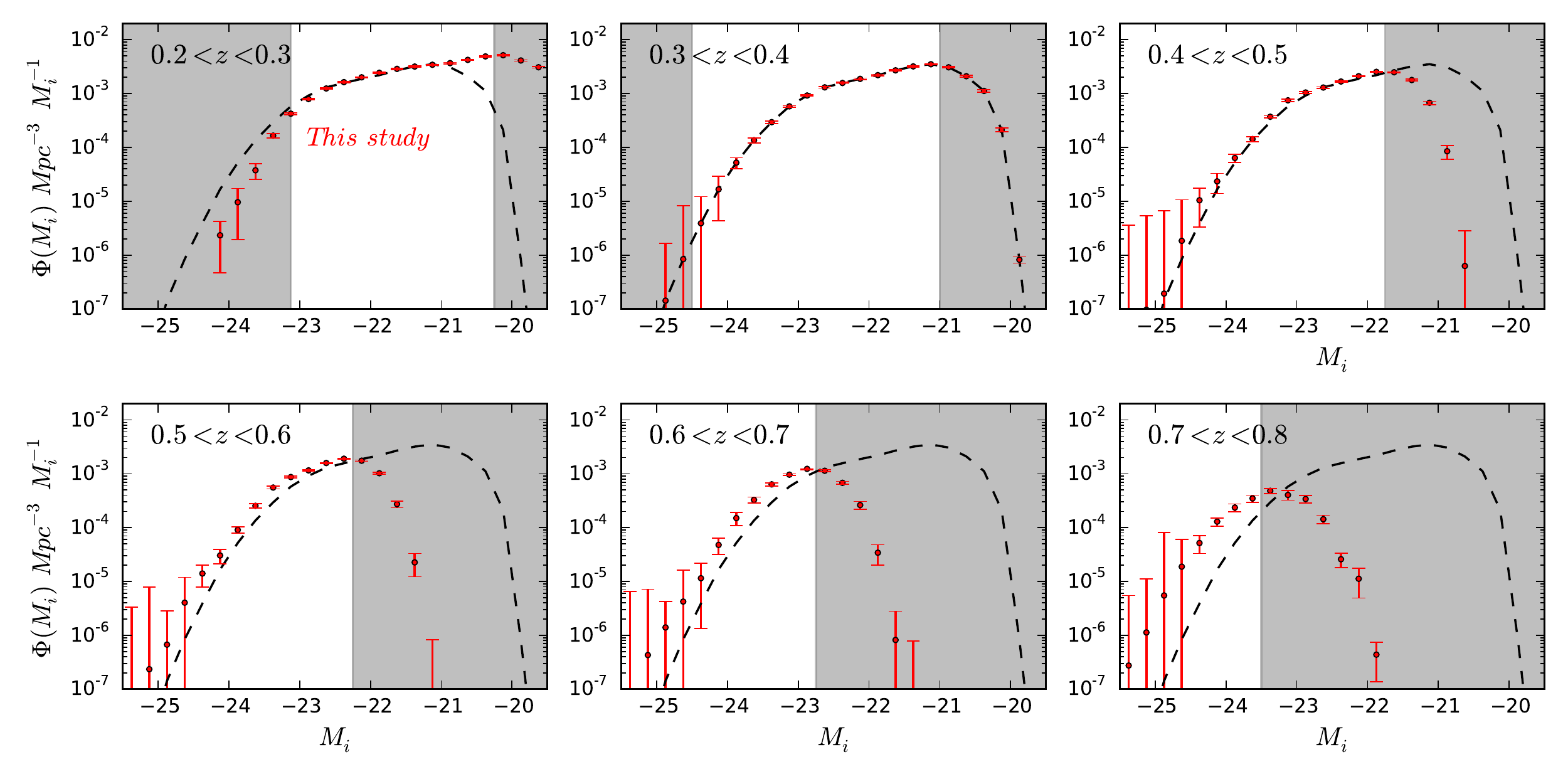}
    \caption{Recovered luminosity functions for real SDSS data in a number of different redshift bins (red). The luminosity function for (0.3 < z < 0.4) is shown as the black dashed line in all redshift bins for reference. Regions where our mass functions are more than 95\% incomplete (due to the photometric sample magnitude cut) are shown in grey.}
    \label{fig:sdss_lf}
\end{figure*}

\textcolor{black}{We now apply the technique to real SDSS data to produce stellar mass and luminosity functions. As described in section \ref{sec:test_clustz_real_data}, we recover redshift distributions of SDSS galaxies, chosen according to many bins of colour and magnitude between $17 < i < 21$. Note that we do not apply any bias correction for the unknown sample since we will test the effect of this later. We compute stellar mass and luminosity distributions for each bin using the colour-mass/luminosity relations of LGalaxies following section \ref{sec:test_mf_lf}.}

\textcolor{black}{Since our reference sample was originally removed from the SDSS sample, we compute stellar masses and luminosities for these galaxies in the same way as for the unknown sample: i.e. for a given colour-redshift bin of our reference sample, we compute the stellar mass or luminosity distribution within the same bin of L-Galaxies. We use spectroscopic redshift distributions instead of clustering redshifts for our reference sample. The agreement between the stellar estimates of the BOSS reference sample using our method with other published estimates is shown in Appendix~\ref{sec:append_2}.}

After adding together the stellar-mass and luminosity distributions from all colour-magnitude bins, we produce global stellar mass and luminosity functions, shown in Figures \ref{fig:sdss_mf} and \ref{fig:sdss_lf}.

\subsection{Galaxy Stellar Mass Functions from SDSS}
\label{sec:mfunc}

The computed stellar mass functions are presented in Figure~\ref{fig:sdss_mf}. The $95\%$ completeness limits are shown in grey, computed as regions where LGalaxies becomes less than $95\%$ complete due to the magnitude cuts. The bright magnitude cut $(i>17)$ is significant in the two lowest redshift bins; however, the impact of this becomes less significant at higher redshifts. The faint magnitude cut becomes more significant at higher redshifts; however, we are still mostly complete at the very high mass end ($\gtrsim 10^{11}$) across the range $(0.4 < z < 0.8)$. Tabulated versions of these mass functions are presented in Appendix \ref{sec:append_3}.

Mass functions for the lowest redshift bins match closely with GAMA mass functions $(z < 0.06)$ over the complete regions, indicating no significant offsets between our masses and GAMA. At the high mass end ($M_{\star}> 10^{11} M_{\odot}$), little evolution is evident over the redshift range $(0.4 < z < 0.8)$, and the mass function is broadly consistent with GAMA ($z<0.06$), implying there is no significant enhancement of the high mass end of the mass function after $z=0.8$. 



\subsection{Luminosity Functions from SDSS}
\label{sec:lfunc}

Our computed luminosity functions are shown in Figure~\ref{fig:sdss_lf}. Magnitudes shown are absolute, dust-corrected magnitudes. Incompleteness is again visible for bright galaxies at low redshifts (due to the $i > 17$ cut); however, beyond redshift 0.4 we are complete for $M_i \lesssim -23.5$, allowing us to compare the evolution of the brightest galaxies across multiple bins. 

\textcolor{black}{There appears to be a significant amount of evolution over the range $(0.3 < z < 0.8)$, with significantly more luminous galaxies present at higher redshifts. If these luminous galaxies are evolving passively, with little ongoing star formation, we would expect their stellar populations to decrease in brightness as young stars die out. \cite{wake08} find similar evolution, and find that this is inconsistent with purely passive evolution. Analysis of these and similar luminosity functions as a test of passive evolution may be of interest for future studies.}

\subsection{The effect of a bias correction on stellar mass functions}
\label{sec:mfunc_b_corr}

\begin{figure*}
\includegraphics[width=\linewidth]{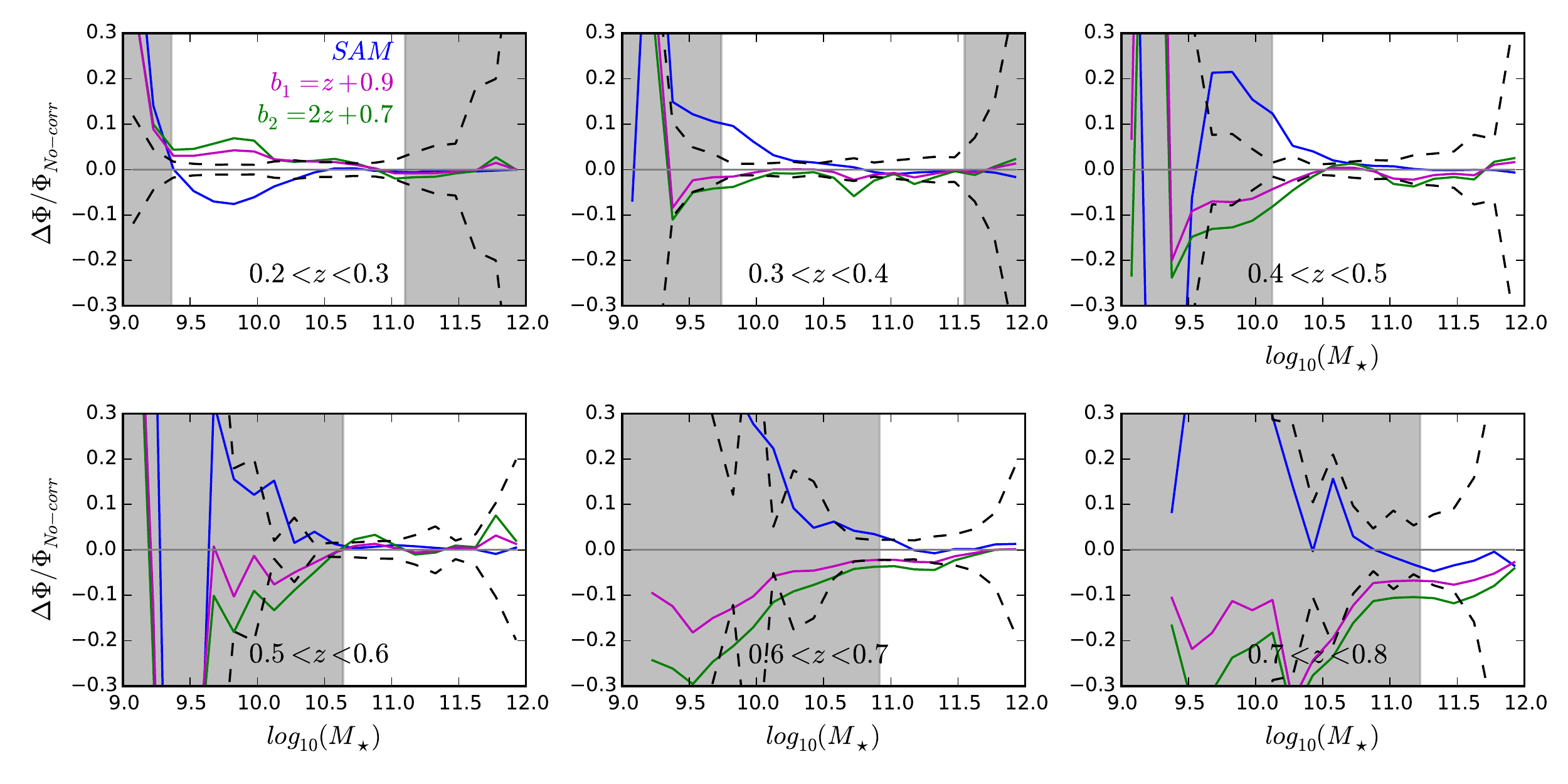}
    \caption{The fractional change in the mass function after applying three different unknown sample bias corrections, shown between $0.2 < z < 0.8$. Our correction from L-Galaxies is shown in blue, and two analytic bias corrections are shown in magenta and green. For reference, the size of the error in the mass function is shown as the black dashed line. Regions where mass functions are more than 95\% incomplete (due to the photometric sample magnitude cut) are shown in grey.}
    \label{fig:mfunc_b_corr}
\end{figure*}

\textcolor{black}{We test how dependant our results are on the choice of unknown sample bias correction as in section~\ref{sec:test_bias_corr}. Figure \ref{fig:mfunc_b_corr} shows the fractional change in the mass function after applying three different unknown sample bias corrections (relative to no correction). We use the correction computed from L-Galaxies in section \ref{sec:clustz_bias_evo_LGal}, and two different analytic bias laws outlined in section \ref{sec:clustz_bias_evo_2}}

\textcolor{black}{Some differences are seen in how the different bias laws affect the mass functions, particularly at lower masses, with the two analytic laws predicting fewer low mass galaxies at high redshifts. At higher masses, however, both the SAM and analytic bias corrections only change the mass function by a few percent, which is normally smaller than, or comparable to the size of our mass function errors. In the analysis of future surveys, where clustering errors will be significantly smaller, the choice of bias correction might play a more significant role. For the data presented here, however, the effect is minimal. When tabulating our mass functions, luminosity and completeness estimates in tables \ref{tab:mass}, \ref{tab:lum}, \ref{tab:comp}, we apply no bias correction, but use the maximum offset in the mass function from the three bias laws an estimate of the systematic error due to the unknown sample bias, which can be added to our errors in quadrature.}

\section{Completeness}
\label{sec:comp}

\begin{figure*}
\includegraphics[width=\linewidth]{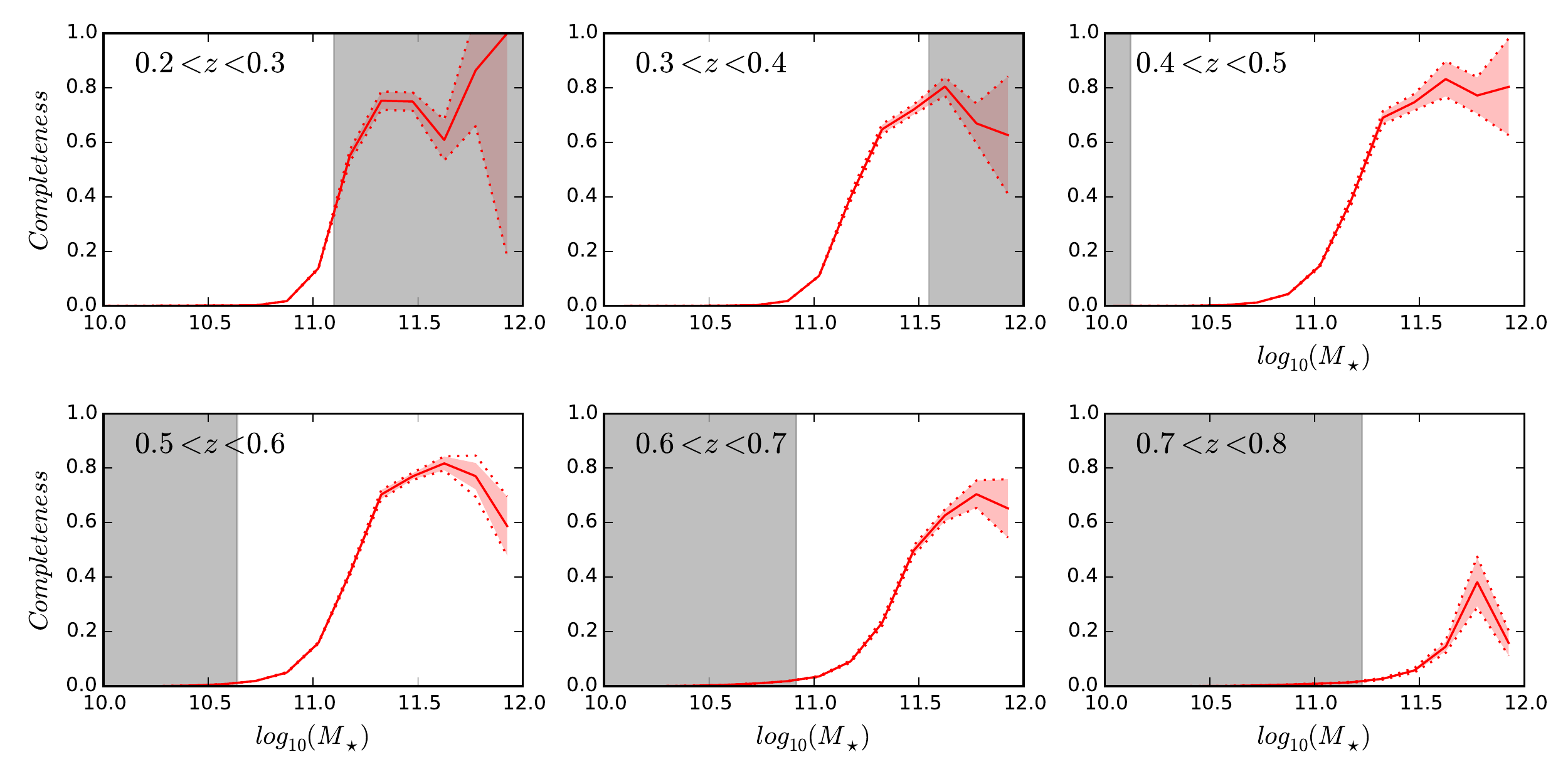}
    \caption{Stellar mass completeness estimates for BOSS between $0.2 < z < 0.8$, computed using the SDSS mass functions recovered in section \ref{sec:mfunc}. Completeness estimates are shown as the solid red line. The shaded red region represents the errors due to the clustering redshifts method, and the dotted red line represents the same error, but with our systematic correction added in quadature. Regions where the mass functions are more than 95\% incomplete (due to the photometric sample magnitude cut) are the grey regions.}
    \label{fig:boss_completeness}
\end{figure*}

Having computed stellar mass functions out to $z=0.8$, we can now measure the stellar mass completeness of the BOSS spectroscopic sample. We first take both the SDSS and BOSS masses computed in section \ref{sec:mfunc}. The completeness at a particular redshift is therefore just the mass function of BOSS at that redshift divided by the SDSS mass function. The resulting completeness is displayed in figure \ref{fig:boss_completeness} for 6 bins of redshift between $0.2 < z < 0.8$. 

At low redshift $z < 0.4$, our SDSS mass functions are not complete at higher masses due to the bright magnitude cut $(i > 17)$. This effect is also true for low masses at higher redshifts due to the faint $(i < 21)$ cut. Completeness estimates of BOSS in these regions may not be fully representative and is shown in grey in figure \ref{fig:boss_completeness}. Between $0.4 < z < 0.8$, however, we are not affected by these cuts over the mass range of BOSS galaxies.

Between $0.2 < z < 0.7$, the stellar mass completeness of BOSS appears similar across all redshifts. Over this redshift range, above $M_{\star} \simeq 10^{11.4} M_{\odot}$, BOSS is roughly $80\%$ complete, with completeness falling to roughly zero at masses lower than $M_{\star} \simeq 10^{11} M_{\odot}$. In the $0.6 < z < 0.7$ bin, incompleteness appears at slightly higher masses than in the lower redshift bins. This decrease in completeness mirrors the decrease in number density of the sample shown in figure \ref{fig:number_dens}, which peaks just above $z=0.5$ and falls off at higher redshifts. Looking in the highest redshift bin, BOSS is around $30\%$ complete, only at the highest masses ($M_{\star} \gtrsim 10^{11.6}$).

Stellar masses are dependent on the the method used to obtain them. When mass functions or completeness measurements are compared between methods, any offsets should be taken in to account. We investigate the difference between our method and different BOSS stellar mass estimates in appendix \ref{sec:append_2}.

\section{Discussion and Conclusions}
\label{sec:conc}

In this study, we have demonstrated that clustering redshifts can be used to successfully recover redshift distributions of galaxies in small bins of colour and magnitude of the SDSS by crosscorrelating with galaxies in the BOSS and eBOSS surveys. The importance of the bias correction becomes significant for fainter galaxies, where photometric errors are large, and galaxies are scattered between colour bins.

\textcolor{black}{We have shown that mass and luminosity functions of mock data can be recovered using these recovered redshift distributions by computing masses using simulations in small bins of colour and redshift. We have also recovered mass functions of real data, and find little evolution at high masses between $0.2 < z < 0.8$, suggesting that the most massive galaxies form most of their mass before this time, and do not evolve significantly in mass afterwards. The lack of evolution over these redshifts agrees well with other studies, for example, \cite{pere08, mous13, leau16, guo18}. In our study, the effect of a bias correction on the recovered mass functions is generally comparable to, or smaller than, the error, however this may not be the case for future large-volume surveys. Our luminosity functions show some evolution with redshift, possibly due to passive evolution.}

We also produce targeting completeness measurements for BOSS using these mass functions, suggesting that over the redshift range $0.2 < z < 0.7$, BOSS is around $80\%$ complete at high masses ($M_{\star} > 10^{11.4} M_{\odot}$), and falling to almost zero below $M_{\star} < 10^{11} M_{\odot}$. In our highest redshift bin $(0.7 < z < 0.8)$ BOSS is strongly affected by incompleteness, and is only about $30\%$ complete at the highest masses $M_{\star} \gtrsim 10^{11.6} M_{\odot}$. We also demonstrate that when comparing mass functions or completeness estimates between methods, significant offsets can be present, which require correction.

\cite{guo18} incorporate an missing fraction (incompleteness) component into their conditional stellar mass function model, and analyse the clustering of BOSS galaxies to produce completeness estimates for BOSS. They find that BOSS is around $80\%$ complete above $M_{\star} \gtrsim 10^{11.3} M_{\odot}$ between $0.2<z<0.6$, with completeness falling off significantly at higher redshifts. This analysis is in good agreement with our results, showing very similar evolution with redshift and mass, although some offsets may be present due to using different mass estimates. \cite{leau16}, discussed in section \ref{sec:intro}, report similar completeness estimates at most redshifts and masses, however  predict close to $100\%$ completeness at the highest masses, which is not shown in \cite{guo18} our estimates.

Ongoing and future large-volume spectroscopic surveys, for example eBOSS, DESI and EUCLID \citep{laur11}, will produce large number of spectra out to higher redshifts. This will firstly allow for better clustering redshifts estimates due to having a larger reference sample, but also produce large spectroscopic galaxy samples, for which incompleteness must be understood. Combining these data with ongoing and future photometric surveys, for example, The Dark Energy Camera Legacy Survey (DECaLS) \citep{dey18}, and The Dark Energy Survey (DES) \citep{des17}, will allow for redshift distributions to be computed out to higher redshifts, and in much smaller bins of colour, due to these new surveys reaching much deeper and having much smaller photometric error.

The methods used in this study, and similar techniques, will therefore be important tools for the next generation of galaxy surveys in order to utilise these large databases, and to understand the galaxy populations present.

\section*{Acknowledgements}

We thank the referee for their careful revision of this manuscript, which has led to important clarifications throughout. 

Funding for the Sloan Digital Sky Survey IV has been provided by the Alfred P. Sloan Foundation, the U.S. Department of Energy Office of Science, and the Participating Institutions. SDSS acknowledges support and resources from the Center for High-Performance Computing at the University of Utah. The SDSS web site is www.sdss.org.




\bibliographystyle{mnras}
\bibliography{bibleography} 

\appendix

\section{Testing the Fitting Scale}
\label{sec:append_1}

\begin{figure*}
	\includegraphics[width=\linewidth]{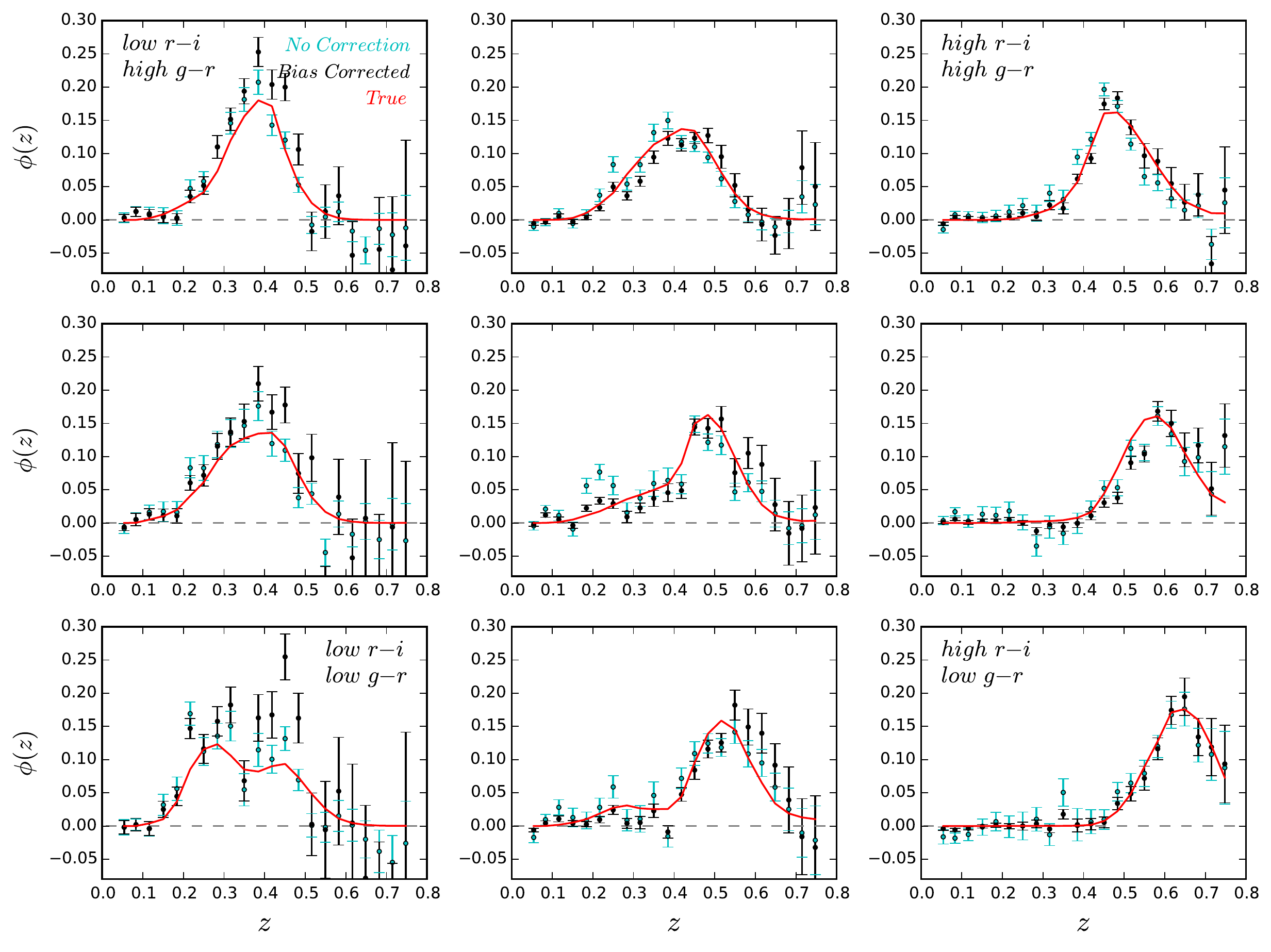}
    \caption{Same as in figure \ref{fig:phi_z_tests}, except the crosscorrelation is integrated over large scales (15 < $r_p$ < 50 Mpc), rather than smaller scales.}
    \label{fig:phi_z_tests_3}
\end{figure*}

Here we show how the choice of fitting scale affects the recovered $\phi(z)$. As in section \ref{sec:test_clustz}, we compute redshift distributions of mock data in small bins of magnitude and colour. Here we show the recovery of several colour bins within the faintest magnitude bin, but rather than integrating the crosscorrelation over small scales, as in figure \ref{fig:phi_z_tests}, we integrate over large scales (15 < $r_p$ < 50 Mpc). The results are shown in figure \ref{fig:phi_z_tests_3}

While redshift distributions are generally recovered successfully, there is a significant amount of extra noise when compared with the small scale recovery (figure \ref{fig:phi_z_tests}). We compute the average error in $\phi(z)$ (i.e. the error due to errors in the correlation functions) for both small scales and large scales. We average this error across all colour bins, and all redshifts; when recovering redshift distributions over large scales, the error is on average 2.4x larger. When noise becomes large, a significant error in normalisation can appear, as seen in, for example, the bin of lowest $r-i$ and highest $g-r$ of figure \ref{fig:phi_z_tests_3}. For this reason, we use only small scale clustering when applying to real data.

\section{Comparing BOSS Mass Functions}
\label{sec:append_2}

\begin{figure*}
	\includegraphics[width=\linewidth]{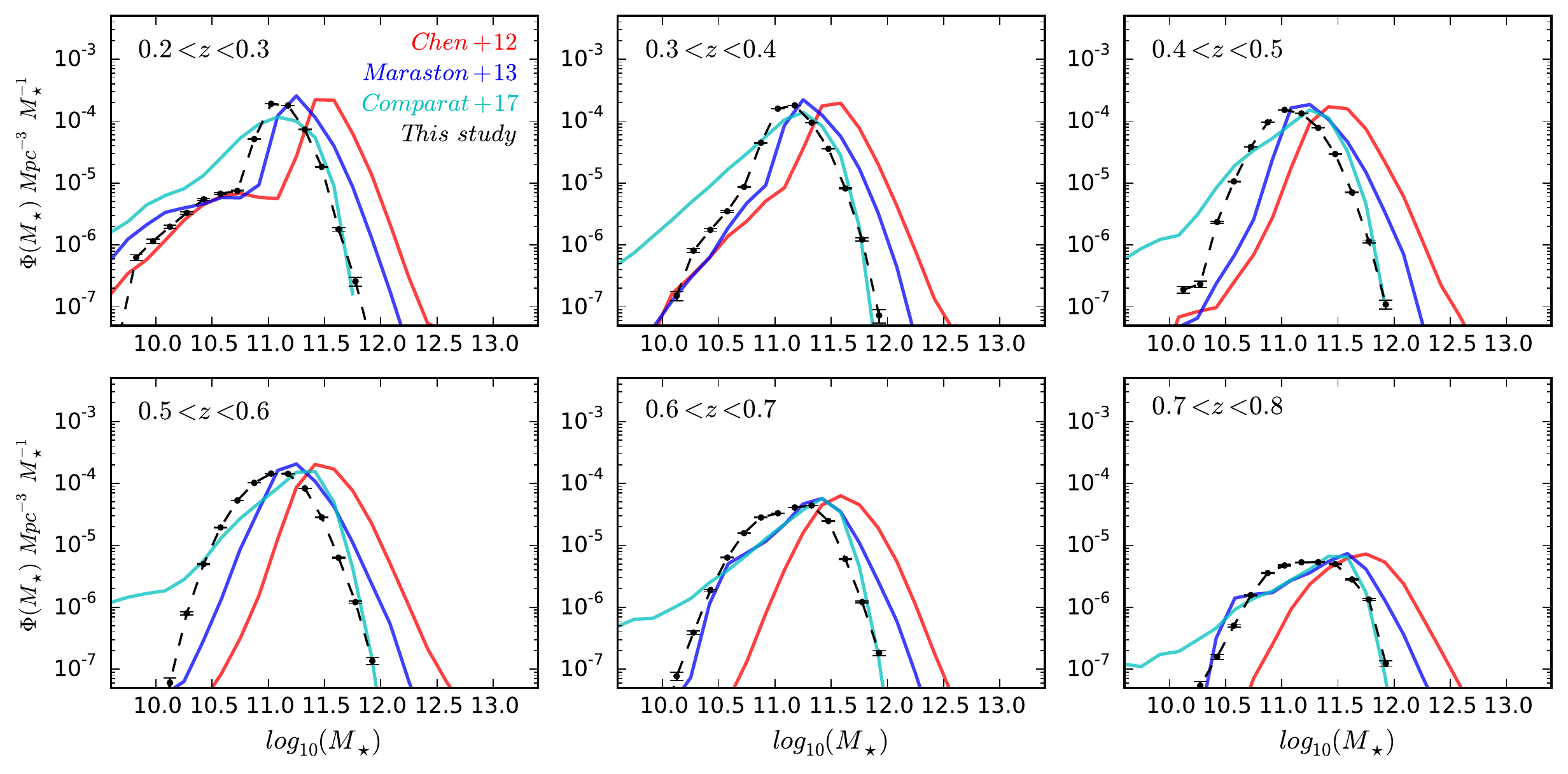}
    \caption{Mass functions of BOSS galaxies using four different methods, shown for six different bins of redshift. Our method described in section \ref{sec:test_mf_lf} is shown in black, along with C12 in red, M13 in blue, and C17 in cyan.}
    \label{fig:boss_mf_comparison}
\end{figure*}

Here we compare, for BOSS galaxies, our mass functions to mass functions computed using three other methods: 1)  \cite{chen12}, hereon Ch12, where galaxy parameters are modeled based on a library of model spectra for which principal components have been identified. 2) \cite{mara13}, hereon Ma13, where stellar population models are fit to the observed $ugriz$ magnitudes, as well as the spectroscopic redshift of each galaxy. 3) \cite{comp17}, hereon Co17, which for a given spectra finds the best-fit combination of single-burst SSPs. All three methods use \cite{mara11} SSPs and a \cite{krou01} IMF. The four mass functions are presented in figure \ref{fig:boss_mf_comparison}.

Although all methods generally agree on the shape of the mass function, there is a clear offset between methods. In particular, Ch12 predicts the highest masses. Both Ma13 and our method predict broadly the same shape as Ch12 at all redshifts, but this is offset towards slightly lower masses. This result may be related to the fact that in our method and Ma13, masses are computed from photometry rather than spectra. The shape of the Co17 mass function appears slightly different. It predicts a larger number of low mass ($M_{\star} < 10^{10} M_{\odot}$) galaxies; however, the number of high mass galaxies is similar to our method. When comparing mass functions or completeness estimates across methods, this offset between different methods must be taken in to account.

\section{Tabulated Results}
\label{sec:append_3}

We present tabulated versions of our stellar mass functions and i-band luminosity functions in tables \ref{tab:mass} and \ref{tab:lum} respectively, and our completeness estimates in table \ref{tab:comp}. In each table, we also present the error in our mass functions due to the clustering redshifts method, and the systematic error due to the bias correction, which can be added together in quadrature.


\begin{table*}
  \label{tab:mass}
  \begin{center}
    \caption{Tabulated stellar mass functions computed as in section \ref{sec:mfunc}}
    \label{tab:table1}
    \begin{tabular}{|cccc|cccc|} 
      \hline
      \multicolumn{4}{c}{0.2 < z < 0.3} & \multicolumn{4}{c}{0.3 < z < 0.4} \\
      \hline
      $\log(M\star)$ $(M_{\odot})$ & $\Phi$ $(10^{-3} Mpc^{-3})$ & $\Phi_{err}$ &$\Phi_{sys}$  &$\log(M\star)$ $(M_{\odot})$ & $\Phi$ $(10^{-3} Mpc^{-3})$ & $\Phi_{err}$  & $\Phi_{sys}$ \\

      9.375 & 8.270 & 0.143 & 0.362 & ... & ... & ... & ...  \\
      9.525 & 6.849 & 0.092 & 0.322& ... & ... & ... & ...  \\
      9.675 & 5.866 & 0.068 & 0.411& ... & ... & ... & ...  \\
      9.825 & 5.557 & 0.067 & 0.420& 9.825 & 4.113& 0.048& 0.394 \\
      9.975 & 5.476 & 0.062 & 0.349& 9.975 & 3.985& 0.048& 0.246\\
      10.125 & 5.321 & 0.099& 0.197& 10.125& 3.946& 0.055& 0.125\\
      10.275 & 5.167 & 0.108& 0.110& 10.275& 4.061& 0.068& 0.078\\
      10.425 & 5.001 & 0.081& 0.097& 10.425& 4.132& 0.051& 0.065\\
      10.575 & 4.555 & 0.074& 0.107& 10.575& 3.896& 0.074& 0.069\\
      10.725 & 3.817 & 0.053& 0.055& 10.725& 3.209& 0.080& 0.187\\
      10.875 & 2.965 & 0.045& 0.008& 10.875& 2.513& 0.039& 0.060\\
      11.025 & 1.395 & 0.030& 0.027& 11.025& 1.479& 0.030& 0.015\\
      ... & ... & ...& ...& 11.175& 0.448& 0.011& 0.014\\
      ... & ... & ...& ...& 11.325& 0.147& 0.004& 0.002\\
      ... & ... & ...& ...& 11.475& 0.048& 0.001& 0.000\\
      \\
      \hline
      
      \multicolumn{4}{c}{0.4 < z < 0.5} & \multicolumn{4}{c}{0.5 < z < 0.6} \\
      \hline
      $\log(M\star)$ $(M_{\odot})$ & $\Phi$ $(10^{-3} Mpc^{-3})$ & $\Phi_{err}$ & $\Phi_{sys}$ & $\log(M\star)$ $(M_{\odot})$ & $\Phi$ $(10^{-3} Mpc^{-3})$ & $\Phi_{err}$ & $\Phi_{sys}$ \\

      10.275 & 2.832   & 0.079  &0.148 & ... & ... & ... &...  \\
      10.425 & 3.391   & 0.035  &0.136 & ... & ... & ... &...  \\
      10.575 & 3.577   & 0.046  &0.072 & ... & ... & ... &...  \\
      10.725 & 3.162   & 0.056  &0.042 & 10.725 & 2.768& 0.045&0.064\\
      10.875 & 2.221   & 0.046  &0.018 & 10.875 & 2.038& 0.038&0.067\\
      11.025 & 1.059   & 0.021  &0.033 & 11.025 & 0.929& 0.018&0.009\\
      11.175 & 0.344   & 0.011  &0.012 & 11.175 & 0.336& 0.011&0.003\\
      11.325 & 0.118   & 0.004  &0.002 & 11.325 & 0.123& 0.006&0.000\\
      11.475 & 0.039   & 0.002  &0.001 & 11.475 & 0.038& 0.001&0.000\\
      11.625 & 0.0092  & 0.0007 &0.0000 & 11.625 & 0.0082& 0.0003&0.0000\\
      11.775 & 0.0014  & 0.0001 &0.0000 & 11.775 & 0.0018& 0.0002&0.0000\\
      11.925 & 0.00025 & 0.00008&0.00000 & 11.925 & 0.00021& 0.00005&0.00000\\
      \\
      \hline
      
      \multicolumn{3}{c}{0.6 < z < 0.7} & \multicolumn{3}{c}{0.7 < z < 0.8} \\
      \hline
      $\log(M\star)$ $(M_{\odot})$ & $\Phi$ $(10^{-3} Mpc^{-3})$ & $\Phi_{err}$ & $\Phi_{sys}$ & $\log(M\star)$ $(M_{\odot})$ & $\Phi$ $(10^{-3} Mpc^{-3})$ & $\Phi_{err}$ & $\Phi_{sys}$ \\

      11.025 & 0.924   & 0.020 &0.068  & ... & ...& ...&...\\
      11.175 & 0.452   & 0.009 &0.056  & ... & ...& ...&...\\
      11.325 & 0.184   & 0.005 &0.033  & 11.325 & 0.178& 0.015&0.018\\
      11.475 & 0.048   & 0.002  &0.019 & 11.475 & 0.073& 0.007&0.008\\
      11.625 & 0.0091  & 0.0004 &0.008 & 11.625 & 0.013& 0.002&0.001\\
      11.775 & 0.0018  & 0.0001 &0.0001 & 11.775 & 0.0033& 0.0012&0.0002\\
      11.925 & 0.00028 & 0.00005&0.00001 & 11.925 & 0.00018& 0.00042&0.00007\\

    \end{tabular}
  \end{center}
\end{table*}


\begin{table*}
  \label{tab:mass}
  \begin{center}
    \caption{Tabulated i-band luminosity functions computed as in section \ref{sec:lfunc}}
    \label{tab:lum}
    \begin{tabular}{|cccc|cccc|} 
      \hline
      \multicolumn{4}{c}{0.2 < z < 0.3} & \multicolumn{4}{c}{0.3 < z < 0.4} \\
      \hline
      $M_i$ $(mag)$ & $\Phi$ $(10^{-3} Mpc^{-3})$ & $\Phi_{err}$ & $\Phi_{sys}$ & $M_i$ $(mag)$ & $\Phi$ $(10^{-3} Mpc^{-3})$ & $\Phi_{err}$ & $\Phi_{sys}$  \\

      ...     & ... & ...&... &      -24.375 & 0.0037& 0.0063&0.0000 \\
      ...     & ... & ...&... &      -24.125 & 0.015& 0.014&0.001 \\
      ...     & ... & ...&... &      -23.875 & 0.053&  0.013&0.004 \\
      ...     & ... & ...&... &      -23.625 & 0.131& 0.013&0.010 \\
      ...     & ... & ...&... &      -23.375 & 0.296& 0.015&0.026 \\
      -23.125 & 0.417 & 0.018 &0.024 & -23.125 & 0.577 & 0.018&0.041\\
      -22.875 & 0.789 & 0.024 &0.036 & -22.875 & 0.907 & 0.024&0.0414    \\
      -22.625 & 1.213 & 0.025 &0.038 & -22.625 & 1.276 & 0.043&0.011    \\
      -22.375 & 1.612 & 0.029 &0.030 & -22.375 & 1.513& 0.043&0.132\\
      -22.125 & 1.972 & 0.037 &0.067 & -22.125 & 1.886& 0.048&0.143 \\
      -21.875 & 2.435 & 0.046 &0.109 & -21.875 & 2.302& 0.049&0.118 \\
      -21.625 & 2.905 & 0.071 &0.122 & -21.625 & 2.804& 0.049&0.149 \\
      -21.375 & 3.221 & 0.078 &0.084 & -21.375 & 3.26& 0.053&0.197 \\
      -21.125 & 3.445 & 0.073 &0.044 & -21.125 & 3.417& 0.107&0.123\\
      -20.875 & 3.778 & 0.077 &0.080 & ...& ...& ...&... \\
      -20.625 & 4.216 & 0.071 &0.173 & ...& ...& ...&... \\
      -20.375 & 4.968 & 0.121 &0.253 & ...& ...& ...&... \\

      \\
      \hline
      
      \multicolumn{4}{c}{0.4 < z < 0.5} & \multicolumn{4}{c}{0.5 < z < 0.6} \\
      \hline
      $M_i$ $(mag)$ & $\Phi$ $(10^{-3} Mpc^{-3})$ & $\Phi_{err}$ & $\Phi_{sys}$ & $M_i$ $(mag)$ & $\Phi$ $(10^{-3} Mpc^{-3})$ & $\Phi_{err}$ & $\Phi_{sys}$  \\

      -25.125 & 0.000 & 0.007&0.000&  -25.125 & 0.000& 0.007&0.000\\
      -24.875 & 0.000 & 0.005&0.000&  -24.875 & 0.001& 0.002&0.000\\
      -24.625 & 0.002 & 0.009&0.000&  -24.625 & 0.005& 0.006&0.001\\
      -24.375 & 0.010 & 0.009&0.001&  -24.375 & 0.015& 0.008&0.002\\
      -24.125 & 0.025 & 0.008&0.002&  -24.125 & 0.029& 0.008&0.002\\
      -23.875 & 0.064 & 0.011&0.007&  -23.875 & 0.089& 0.012&0.004\\
      -23.625 & 0.147 & 0.014&0.015&  -23.625 & 0.254& 0.024&0.037\\
      -23.375 & 0.383 & 0.018&0.024&  -23.375 & 0.534& 0.029&0.036\\
      -23.125 & 0.735 & 0.035&0.024& -23.125 & 0.864& 0.025&0.037\\
      -22.875 & 1.053 & 0.040&0.023& -22.875 & 1.174& 0.025&0.049\\
      -22.625 & 1.306 & 0.031&0.064& -22.625 & 1.567& 0.022&0.065\\
      -22.375 & 1.677 & 0.030&0.081& -22.375 & 1.842& 0.028&0.082\\
      -22.125 & 2.132 & 0.029&0.133& .. & ..& ...&...\\
      -21.875 & 2.490 & 0.045&0.160&... & ...& ...&...\\
      -21.625 & 2.437 & 0.078&0.144& ... & ...& ...&...\\
      \\
      \hline
      
      \multicolumn{4}{c}{0.6 < z < 0.7} & \multicolumn{4}{c}{0.7 < z < 0.8} \\
      \hline
      $M_i$ $(mag)$ & $\Phi$ $(10^{-3} Mpc^{-3})$ & $\Phi_{err}$ & $\Phi_{sys}$ & $M_i$ $(mag)$ & $\Phi$ $(10^{-3} Mpc^{-3})$ & $\Phi_{err}$ & $\Phi_{sys}$  \\

      -25.125 & 0.000 & 0.007&0.000&  -25.125 & 0.001& 0.014&0.000\\
      -24.875 & 0.001 & 0.001&0.000&  -24.875 & 0.005& 0.015&0.000\\
      -24.625 & 0.004 & 0.013&0.000&  -24.625 & 0.018& 0.024&0.000\\
      -24.375 & 0.011 & 0.009&0.001&  -24.375 & 0.033& 0.018&0.000\\
      -24.125 & 0.044 & 0.014&0.001&  -24.125 & 0.124& 0.026&0.001\\
      -23.875 & 0.148 & 0.014&0.007&  -23.875 & 0.229& 0.023&0.007\\
      -23.625 & 0.341 & 0.036&0.016&  -23.625 & 0.388& 0.036&0.016\\
      -23.375 & 0.579 & 0.028&0.025&  ... & ...& ... &...\\
      -23.125 & 0.949 & 0.025&0.054&  ... & ...& ... &...\\
      -22.875 & 1.243 & 0.050&0.085&  ... & ...& ... &...\\
      \\

    \end{tabular}
  \end{center}
\end{table*}

\begin{table*}
  \label{tab:comp}
  \begin{center}
    \caption{Tabulated stellar mass completeness for BOSS computed as in section \ref{sec:comp}}
    \label{tab:comp}
    \begin{tabular}{|cccc|cccc|} 
      \hline
      \multicolumn{4}{c}{0.2 < z < 0.3} & \multicolumn{4}{c}{0.3 < z < 0.4} \\
      \hline
      $\log(M\star)$ $(M_{\odot})$ & $Completeness$ & $\sigma_{comp}$ & $\sigma_{sys}$ & $\log(M\star)$ $(M_{\odot})$ & $Completeness$ & $\sigma_{comp}$ & $\sigma_{sys}$\\

      10.575 & 0.0015 & 0.0001&0.0000 &10.575& 0.0009& 0.0001&0.0000 \\
      10.725 & 0.0020 & 0.0001&0.0000& 10.725& 0.0027& 0.0001&0.0001 \\
      10.875 & 0.0178 & 0.0003&0.0000& 10.875& 0.0180& 0.0003&0.0002 \\
      11.025 & 0.1380 & 0.0028&0.0011& 11.025& 0.1104& 0.0016&0.0004 \\
      ... & ... & ...& ...&       11.175& 0.3997& 0.0085&0.0068 \\
      ... & ... & ...& ...&       11.325& 0.6478& 0.0158&0.0035 \\
      ... & ... & ...& ...&       11.475& 0.7198& 0.0183&0.0022 \\
      \\
      \hline
      
      \multicolumn{4}{c}{0.4 < z < 0.5} & \multicolumn{4}{c}{0.5 < z < 0.6} \\
      \hline
      $\log(M\star)$ $(M_{\odot})$ & $Completeness$ & $\sigma_{comp}$ & $\sigma_{sys}$ & $\log(M\star)$ $(M_{\odot})$ & $Completeness$ & $\sigma_{comp}$ & $\sigma_{sys}$\\

      10.575 & 0.0031   & 0.0001   &0.0001& ... & ... & ...   &...\\
      10.725 & 0.0124   & 0.0002   &0.0001& 10.725 & 0.0190& 0.0004&0.0002\\
      10.875 & 0.0438   & 0.0009   &0.0001& 10.875 & 0.0501& 0.0010&0.0009\\
      11.025 & 0.1473   & 0.0025   &0.0028& 11.025 & 0.1586& 0.0024&0.0005\\
      11.175 & 0.3893   & 0.0095   &0.0078& 11.175 & 0.4146& 0.0095&0.0009\\
      11.325 & 0.6904   & 0.0204   &0.0044& 11.325 & 0.7030& 0.0162&0.0005\\
      11.475 & 0.7469   & 0.0281   &0.0025& 11.475 & 0.7699& 0.0111&0.0021\\
      11.625 & 0.8322  & 0.0637  &0.0025& 11.625 & 0.8172& 0.0251&0.0005\\
      11.775 & 0.7719  & 0.0660  &0.0013& 11.775 & 0.7703& 0.0484&0.0274\\
      11.925 & 0.8036 & 0.1761 &0.0012& 11.925 & 0.5874& 0.1085&0.00066\\
      \\
      \hline
      
      \multicolumn{4}{c}{0.6 < z < 0.7} & \multicolumn{4}{c}{0.7 < z < 0.8} \\
      \hline
      $\log(M\star)$ $(M_{\odot})$ & $Completeness$ & $\sigma_{comp}$ & $\sigma_{sys}$ & $\log(M\star)$ $(M_{\odot})$ & $Completeness$ & $\sigma_{comp}$ & $\sigma_{sys}$\\

      11.025 & 0.0357   & 0.0007 &0.0007  & ... & ...& ... & ...\\
      11.175 & 0.0909   & 0.0018 &0.0025  & ... & ...& ... & ...\\
      11.325 & 0.2318   & 0.0061 &0.0058  & 11.325 & 0.0270& 0.0019 &0.0016\\
      11.475 & 0.4971   & 0.0137 &0.0042  & 11.475 & 0.0576& 0.0046 &0.0038\\
      11.625 & 0.6266  & 0.0212 &0.0011  & 11.625 & 0.1451& 0.0177 &0.0054\\
      11.775 & 0.7042  & 0.0499 &0.0007  & 11.775 & 0.3818& 0.0888 &0.0050\\
      11.925 & 0.6521 & 0.1070 &0.0003 & 11.925 & 0.1583& 0.0458 &0.0005\\

    \end{tabular}
  \end{center}
\end{table*}


\bsp	
\label{lastpage}
\end{document}